\documentclass[manuscript]{acmart}

\AtBeginDocument{%
  }

\setcopyright{acmlicensed}
\acmJournal{TOSEM}

\usepackage{multirow}
\usepackage{xspace}
\usepackage{tabularx}
\usepackage{tablefootnote}
\usepackage[most]{tcolorbox}
\usepackage[version=4]{mhchem}
\usepackage{xcolor}
\tcbuselibrary{skins}
\usepackage{color}
\usepackage{tabularray}
\usepackage{array}
\usepackage{booktabs}  
\usepackage{resizegather}
\usepackage{enumitem}

\begin{document}

\title{A Goal-Driven Survey on Root Cause Analysis}

\newcommand{\xl}[1]{\textcolor{blue}{XL: #1}}
\newcommand{\ie}{{\em i.e.},\xspace}
\newcommand{\eg}{{\em e.g.},\xspace}
\newcommand{\vs}{\textit{vs.}\xspace}

\author{Aoyang Fang}
\email{aoyangfang@link.cuhk.edu.cn}
\author{Haowen Yang}
\email{222010523@link.cuhk.edu.cn}
\author{Haoze Dong}
\email{haozedong@link.cuhk.edu.cn}
\author{Qisheng Lu}
\email{qishenglu@link.cuhk.edu.cn}
\author{Junjielong Xu}
\email{junjielongxu@link.cuhk.edu.cn}
\author{Pinjia He}
\email{hepinjia@cuhk.edu.cn}
\affiliation{%
  \institution{The Chinese University of Hong Kong, Shenzhen}
  \city{Shenzhen}
  \state{Guangdong}
  \country{China}
}

\renewcommand{\shortauthors}{Fang et al.}

\begin{abstract}
Root Cause Analysis (RCA) is a crucial aspect of incident management in large-scale cloud services.
While the term \textit{root cause analysis} or \textit{RCA} has been widely used, different studies formulate the task differently.
This is because the term "RCA" implicitly covers tasks with distinct underlying goals. 
For instance, the goal of localizing a faulty service for rapid triage is fundamentally different from identifying a specific functional bug for a definitive fix. 
However, previous surveys have largely overlooked these goal-based distinctions, conventionally categorizing papers by input data types (\eg metric-based vs. trace-based methods). 
This leads to the grouping of works with disparate objectives, thereby obscuring the true progress and gaps in the field.
Meanwhile, the typical audience of an RCA survey is either laymen who want to know the goals and big picture of the task or RCA researchers who want to figure out past research under the same task formulation.
Thus, an RCA survey that organizes the related papers according to their goals is in high demand.
To this end, this paper presents a goal-driven framework that effectively categorizes and integrates \totalrcapaper papers on RCA in the context of cloud incident management based on their diverse goals, spanning the period from 2014 to 2025.
In addition to the goal-driven categorization, it discusses the ultimate goal of all RCA papers as an umbrella covering different RCA formulations.
Moreover, the paper discusses open challenges and future directions in RCA.
\end{abstract}
\newcommand{\totalrcapaper}{135\xspace}
\newtcolorbox[auto counter, number within=section]{takeaway}[2][]{%
  colback=black!10!white, colframe=black!20!black, fonttitle=\bfseries, 
  title=#2~\thetcbcounter: #1, rounded corners}

\newcommand{\etal}{et al.}



\received{20 February 2007}
\received[revised]{12 March 2009}
\received[accepted]{5 June 2009}

\maketitle

\section{Introduction}
\label{sec:introduction}

Microservices have emerged as the favored architecture for cloud-native development in the era of cloud computing.
The adoption of microservice architecture can break down large, monolithic software applications into numerous smaller, more manageable components. 
This division facilitates parallel development across various software segments and enhances overall agility and efficiency.
However, unlike monolithic applications, where components are tightly integrated and easier to trace, microservices operate as separate entities that interact through well-defined interfaces. 
This increases the complexity of the interactions between services, making it more challenging to pinpoint the origin of incidents. 
Incidents in large-scale cloud services can lead to significant financial losses and disruptions in critical services~\cite{openai2024incident,azure_status,awsincident}. 
For example, the 2024 CrowdStrike-related IT outages~\cite{crowdstrike2024} caused widespread service outages affecting Azure, Teams, and Xbox Live, resulting in prolonged downtime for many users, and the worldwide financial damage has been estimated to be at least 10 billion US dollars.
Root Cause Analysis (RCA) has emerged as a critical phase in identifying the underlying reasons for incidents. 
Traditional RCA requires substantial human effort to sift through vast quantities of telemetry data, code, and other resources~\cite{google_sre_effective_troubleshooting}.
Operators involved are expected to have extensive domain knowledge, such as a comprehensive understanding of the operational environment, familiarity with the codebase, and in cloud infrastructure scenarios, even insights into operating system-level components like the Linux kernel.
Due to the complex inter-service dependency and intra-service business logic, it is challenging for operators to fully understand the full scope of service interactions~\cite{google_sre_workbook_incident_response}, thus making root cause analysis difficult.

The practical application of RCA is not a monolithic task. 
As detailed in our formalization (Section~\ref{sec:preliminaries}), its objectives are fundamentally shaped by the incident management lifecycle (Section~\ref{sec:background}). 
This distinction is best understood by differentiating two key concepts, which we will formalize in Section~\ref{sec:preliminaries}.
A Site Reliability Engineer (SRE) focused on rapid mitigation (minimizing Mean-Time-To-Recovery) seeks to identify the \textit{trigger}: an event, such as a sudden traffic spike, that activates a latent flaw. 
In contrast, a developer aiming for a permanent fix must find the underlying \textit{root cause}: the fundamental flaw itself, such as a buggy code commit.
This difference in objectives, compounded by the varying data access permissions across roles, creates fundamentally different requirements for input data, analysis depth, and output granularity.
This inherent complexity has led to significant fragmentation in RCA research. 
Most studies implicitly tailor their problem formulation to a specific goal, creating tight coupling between methods and narrow task definitions. 
For example, one line of research focuses on identifying faulty services from metrics~\cite{li2022causal,pham2024baro,wang2023incremental} (typical SRE goals), while another aims to pinpoint buggy code changes from traces and logs~\cite{yu2023nezha,gu2023trinityrcl} (developer goals). 
This has resulted in a combinatorial explosion of task-specific solutions, making it exceedingly difficult to compare methods, generalize findings, or build a cohesive understanding of the field.

Consequently, the conventional survey's categorization based on input data types (\eg logs, traces, metrics)~\cite{soldani2022anomaly,zhang2024failure,wang2024comprehensive} fails to capture the underlying objectives that drive these formulations. 
This taxonomy is misleading because the relationship between input data and research objectives is not one-to-one; it arbitrarily groups studies with different goals that happen to use similar data, while separating those with shared goals that use different data types.
For newcomers or practitioners seeking a high-level understanding, this view obscures the bigger picture of what RCA can achieve. 
For researchers, it makes comparing the true capabilities of different methods nearly impossible, as a technique's performance is inextricably tied to a narrow, implicit goal.
This lack of a unified, goal-oriented perspective imposes a significant cognitive load, impeding not only the synthesis of academic knowledge but also the integration of disparate tools into a cohesive industrial workflow.

To this end, we anchor our perspective in the overarching goal of incident management: minimizing the \textbf{Mean Time to Recovery} (MTTR). 
As discussed in Section~\ref{sec:background}, this lifecycle involves a chain of activities where the output of one stage, with a specific granularity, becomes the input for the next. 
The effectiveness of RCA is central to this process, as its output directly dictates the speed and precision of resolution. 
A coarse-grained root cause may only permit service rollbacks, whereas a fine-grained cause, such as a specific faulty code commit, enables targeted fixes and prevents future recurrence.
This lifecycle-centric view reveals that an ideal RCA solution must satisfy a set of fundamental requirements to be effective in practice.

Instead of categorizing research by input data types, we argue that a more insightful taxonomy should be based on the inherent goals an RCA system must achieve to excel at each stage of this lifecycle. 
We derive these goals by analyzing the key challenges from four perspectives: the data foundation (Input), the analytical core (Inference), the usability of results (Output), and practical deployment constraints (Efficiency).
An ideal RCA system must be able to:
correlate \textbf{multi-dimensional data},
be \textbf{robust} to imperfect data,
\textbf{adaptively learn} from system changes,
provide \textbf{interpretable} and \textbf{multi-granularity} results that are \textbf{actionable},
all while maintaining \textbf{real-time performance}.

These seven goals, which include multi-dimensional data correlation, robustness, adaptive learning, real-time performance, interpretability, multi-granularity, and actionability, form the pillars of our goal-driven survey.
They are not arbitrary but are directly derived from the practical needs of the incident management lifecycle, and will be formally defined in Section~\ref{sec:preliminaries}.

To systematically investigate the field through this goal-driven lens, we first establish a formal framework for the ideal RCA problem, defined as a function $\mathcal{F}: \mathcal{O} \rightarrow \mathcal{G}$ that maps rich observational data ($\mathcal{O}$) to a complete incident propagation graph ($\mathcal{G}$) (Section~\ref{sec:preliminaries}).
This framework serves as a "north star," allowing us to deconstruct and unify the disparate problem formulations found in the literature.
Following a rigorous survey methodology (Section~\ref{sec:surveymethod}), we collected and analyzed \totalrcapaper papers from top-tier venues.
Each paper is mapped onto our seven-goal taxonomy, enabling a structured analysis of the field's state of the art.
Beyond this categorization, we provide a comprehensive overview of the research landscape, including publication trends, a summary of public benchmarks, datasets, and open-source tools (Section~\ref{sec:trend}).
Finally, we discuss the implications of our framework, identify critical gaps between current research and the ideal RCA, and outline promising future research frontiers (Section~\ref{sec:discussion}).

This work makes the following contributions:

\begin{itemize}
    \item \textbf{A Formal Framework and Goal-Driven Taxonomy.} We propose a formal definition of the ideal RCA problem and introduce a novel, goal-driven taxonomy based on seven fundamental objectives. This framework moves beyond superficial data-based categorizations to provide a more insightful lens for understanding and comparing RCA research.
    \item \textbf{A Systematic and Comprehensive Survey.} We conduct a systematic review of \totalrcapaper papers, analyzing each through our goal-driven taxonomy. This provides a structured overview of the state of the art, revealing the underlying design trade-offs and evolutionary trends in the field.
    \item \textbf{A Curated Overview of the Research Landscape.} We compile and analyze the distribution of research focus, public benchmarks, datasets, and open-source tools. This serves as a valuable guide for researchers and practitioners, while also highlighting critical limitations in existing resources, such as the lack of ground-truth propagation graphs.
    \item \textbf{Identification of Gaps and Future Frontiers.} We identify the significant gap between the current "point-finding" paradigm and the ideal "graph-building" RCA. Based on this analysis, we outline key future research directions, including the need for next-generation benchmarks and unified models for causal graph generation.
\end{itemize}

\begin{figure}[t]
    \centering
    \includegraphics[width=1\linewidth]{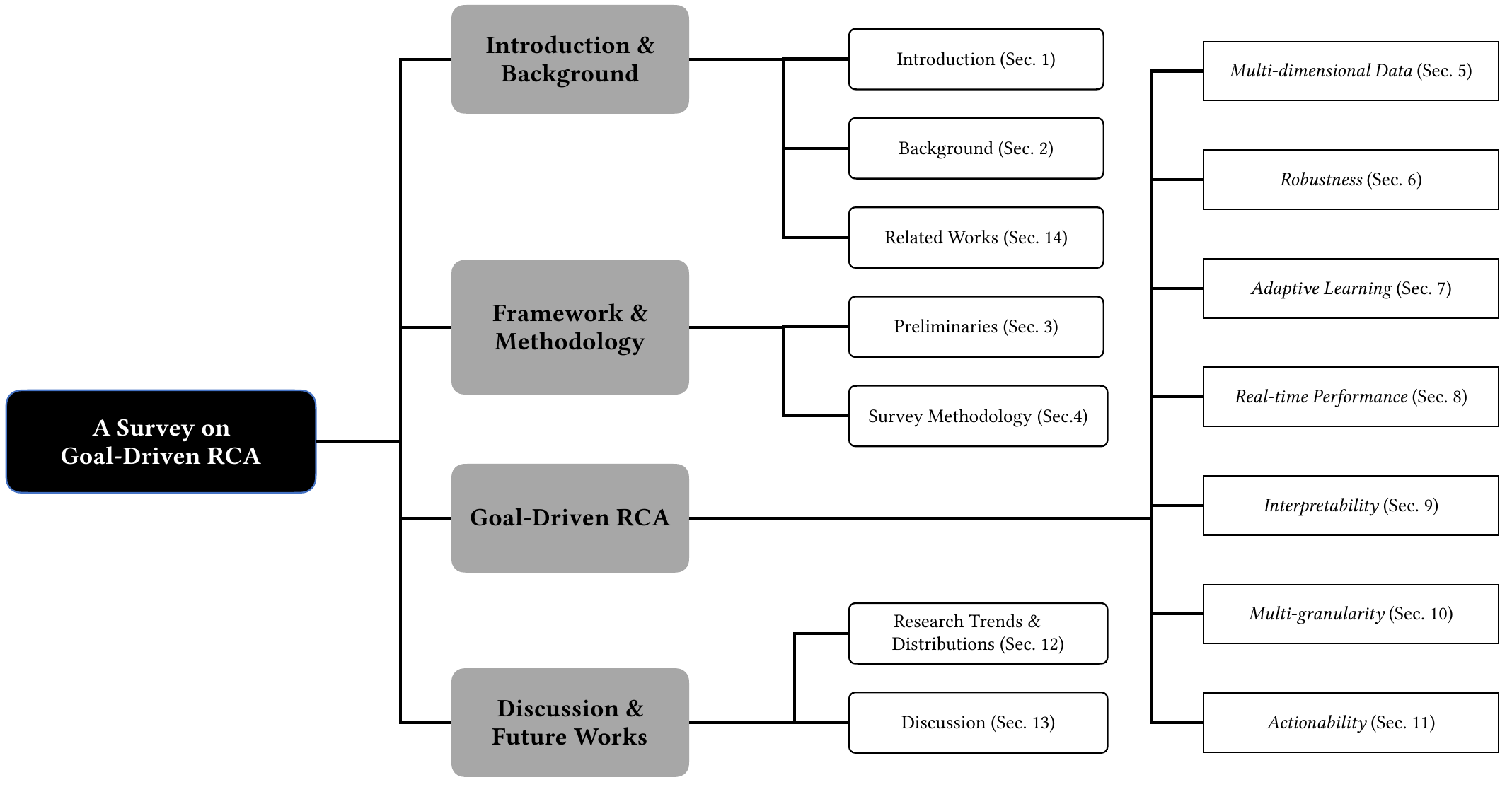}
    \caption{The structure of this survey. We first introduce the background, our formal framework, and survey methodology (\S\ref{sec:background}-\S\ref{sec:surveymethod}). Then, we systematically analyze RCA research through the lens of a seven-goal taxonomy (\S\ref{sec:multidimension}-\S\ref{sec:goal7_actionability}). Finally, we discuss research trends, future opportunities, and related work before concluding (\S\ref{sec:trend}-\S\ref{sec:conclusion}).}
    \label{fig:structure}
\end{figure}

The remainder of this paper is structured as follows, also illustrated in Fig.~\ref{fig:structure}.
Section~\ref{sec:background} and Section~\ref{sec:preliminaries} introduce the background and our formal framework.
Section~\ref{sec:surveymethod} details our survey methodology.
Section~\ref{sec:multidimension} to Section~\ref{sec:goal7_actionability} present a systematic analysis of RCA research through the lens of our seven-goal taxonomy.
Section~\ref{sec:trend} and Section~\ref{sec:discussion} discuss research trends and future opportunities.
Finally, Section~\ref{sec:relatedwork} reviews related surveys, and Section~\ref{sec:conclusion} concludes the paper.

\section{BACKGROUND}
\label{sec:background}
\subsection{MICROSERVICE}

\paragraph{Core Concepts} The microservice is an \textit{implementation of decomposition philosophy} in software architecture, which advocates breaking down complex systems into smaller, independent components.
Specifically, the microservice architecture divides a single application into a collection of small, lightweight services, each running in its own process and communicating via lightweight methods, often using an HTTP API. 
It emphasizes agile, DevOps practices, decentralized data management, and governance \cite{di2017research}.

\paragraph{Technology Stack} Microservice architecture is supported by a series of infrastructure systems and techniques that work together seamlessly. 
Firstly, the development of microservices begins with development frameworks like Spring Boot~\cite{springboot} and Dubbo~\cite{dubbo}, which facilitate the creation of microservices by providing essential functionalities such as REST clients, database integration, externalized configuration, and caching. 
Subsequently, these microservices are deployed using containerization tools like Docker~\cite{docker}, which enhance portability, flexibility, efficiency, and speed. 
To manage these containers effectively, runtime infrastructure frameworks such as Spring Cloud~\cite{springcloud}, Mesos~\cite{mesos}, Kubernetes~\cite{kubernetes}, and Docker Swarm~\cite{docker} are employed, offering capabilities like configuration management, service discovery, service registry, and load balancing. 
Finally, to ensure efficient and reliable development and deployment processes, continuous integration and delivery tools such as Jenkins~\cite{jenkins} and GitLab CI/CD~\cite{gitlab} are utilized to support ongoing integration and delivery efforts. 

\paragraph{Benefits of Microservice Architecture} Microservice architecture offers several notable benefits, making it a popular choice for modern application development. 
By allowing each service to be updated independently, it ensures that changes or failures in one service do not impact the entire application. 
Additionally, it supports independent scaling, which enhances system flexibility and optimizes resource utilization. 
This architecture also facilitates parallel development, enabling multiple teams to work on different services simultaneously, thereby accelerating development speed \cite{Microservices14}.

\paragraph{Challenges and Complexities} However, as systems transition from monolithic to microservice architectures, complexity shifts from internal code to interactions between services. 
This shift introduces new challenges, particularly in monitoring and troubleshooting, as the interactions between services become more intricate.
Modern applications, typically involving hundreds of interconnected services, significantly complicate monitoring efforts~\cite{wu2021identifying}. 
For instance, distinguishing between individual service failures and cascading effects from other service failures becomes challenging. 
Moreover, many microservice failures originate from external environments, like their runtime environments, communication, or coordination issues. 

\subsection{INCIDENT MANAGEMENT}

\begin{figure}
    \centering
    \includegraphics[width=0.8\linewidth]{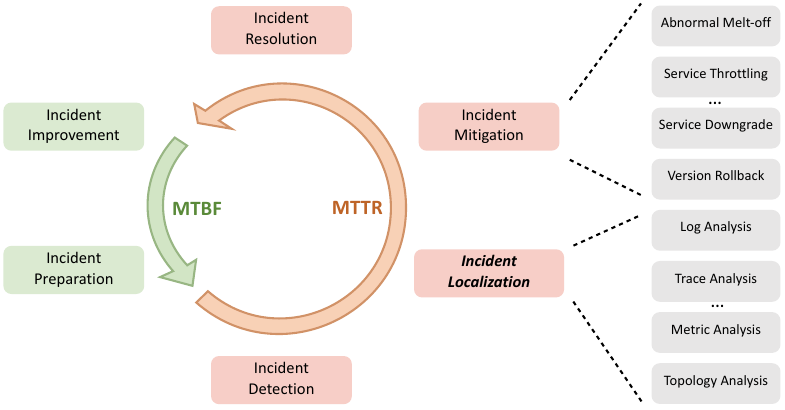}
    \caption{Procedure of incident management. Starting from \textbf{incident preparation}, which involves techniques like software testing, canary releases, and disaster recovery simulations to closely mimic real-world conditions and prepare for potential issues. Once the product is deployed, it enters the \textbf{incident detection} stage, where monitoring systems, anomaly detection, and customer reports are used to identify any abnormalities in runtime services.
    If an issue arises, the process moves to the \textbf{incident localization} phase. Here, SREs work to quickly determine which service is the root cause and further pinpoint the specific underlying issues, including analyzing the log, metric, trace, etc.
    To halt the failure's propagation, the next step is \textbf{incident mitigation}, where common actions might include downgrading the service or rolling back to a previous code version, depending on the identified root cause.
    Following mitigation, the focus shifts to \textbf{incident resolution}. 
    During this phase, SREs and developers collaborate to resolve the incident completely. 
    Once the system is fully restored, the final phase is a \textbf{incident improvement} (postmortem review), which involves analyzing the incident to extract lessons learned and implementing measures to prevent future occurrences.}
    \label{fig:incidentmanagement}
\end{figure}

Drawing on Google's incident management practices~\cite{googlesre}, as illustrated in Fig.\ref{fig:incidentmanagement}, we can break down the process into several key stages. 
This lifecycle begins with proactive \emph{preparation} and moves into reactive phases like \emph{detection}, \emph{localization}, \emph{mitigation}, and \emph{resolution} once an incident occurs, finally concluding with \emph{improvement} through postmortems.

\paragraph{The Pivotal Role of RCA in Incident Response}
Within this lifecycle, Root Cause Analysis (RCA) is not a single, isolated step but a continuous and multi-faceted process that bridges incident detection with effective mitigation and resolution. 
In a practical corporate setting, the nature of RCA dynamically adapts based on the incident's progression and the immediate goal.
These practical demands directly motivate the fundamental goals that an ideal RCA system must pursue, which we will formalize in Section~\ref{sec:preliminaries}.

The RCA process typically begins immediately after an incident is detected. 
The initial phase is often a coarse-grained analysis, commonly known as \emph{triage}. 
The primary goal here is not to find the precise bug, but to quickly identify the responsible service or component and route the alert to the correct on-call team~\cite{chen2019empirical,ghosh2022fight}. 
This demand for speed highlights the critical need for \emph{real-time performance} in any practical RCA tool.
For example, a high-level alert on transaction failures might trigger a triage process that, by examining system topology and top-level service metrics, determines the payment service is the most likely culprit, thus assigning the incident to the payment team.

Once the incident is assigned, the responsible team, typically Site Reliability Engineers (SREs), performs a more in-depth RCA with the immediate goal of \emph{mitigation}. 
At this stage, the objective is to find a "good enough" root cause to stop the bleeding (i.e., minimize Mean Time to Recovery, MTTR). 
SREs must correlate heterogeneous data sources like metrics, logs, traces, and recent changes (\eg deployments, feature flag toggles) to localize the fault.
This process underscores the challenge of achieving effective \emph{multi-dimensional data correlation}.
Furthermore, since this data is often incomplete or noisy, the underlying methods must exhibit \emph{robustness}.
The output of this RCA directly enables mitigation actions, such as service rollbacks, emphasizing the need for findings to be \emph{actionable}. 
For instance, identifying a specific canary instance as faulty leads to its removal from the load balancer.

After the service is stabilized through mitigation, the focus shifts to \emph{permanent resolution}. 
This involves an even deeper and more fine-grained RCA, often conducted by developers in collaboration with SREs. 
The goal now is to uncover the precise, underlying bug or misconfiguration, which requires the RCA results to offer \emph{multi-granularity} views, from service-level down to code-level.
This requires meticulous forensic analysis, potentially down to a specific line of code or a configuration value. 
To be useful for developers, the causal chain leading to the failure must be clear, demanding a high degree of \emph{interpretability} from the RCA model.
For example, while the SRE's RCA identified a faulty deployment for rollback, the developer's RCA must pinpoint the exact code commit that introduced a memory leak. 
This fine-grained output is essential for developing a permanent fix, ensuring the incident does not recur and allowing the system to \emph{adaptively learn} from past failures.

This progression illustrates that RCA's required input and desired output granularity evolve throughout the incident lifecycle. 
It begins broadly (which team is responsible?), narrows for mitigation (which service or deployment should be rolled back?), and becomes highly specific for resolution (which line of code/config needs to be fixed?). 
The effectiveness of the entire incident management process hinges on the ability to perform RCA at these varying levels of depth and speed.

\section{PRELIMINARIES}\label{sec:preliminaries}

This section introduces the key terminology and definitions that form the foundation of this paper.
First, we introduce the core challenges and fundamental trade-offs inherent in RCA, from which we derive the seven goals that guide our survey.
We then present a formal definition of the RCA problem to provide a unified conceptual model for analyzing existing work.
Understanding these concepts is essential for discussing the challenges and goals of root cause analysis.

\subsection{The Core Challenge: The Effectiveness-Data-Cost Triangle} \label{sec:challenges}

Achieving the ideal goals of RCA is profoundly difficult.
The challenges stem from the inherent complexity of telemetry data and a systemic trade-off that governs all practical RCA deployments.
This tension forces strategic compromises that shape how existing research approaches the RCA problem.

\paragraph{Data Complexity: Fragmented, Imperfect, and Overwhelming Telemetry}
The observation space $\mathcal{O}$, which we will formally define in Section~\ref{sec:formal_rca}, presents formidable obstacles to effective analysis.
Telemetry data is inherently \textit{fragmented}, with different data types (\eg logs, metrics, traces) offering complementary but siloed views of system operation.
The data is also \textit{imperfect}, suffering from inaccuracies due to measurement errors, incompleteness from collection failures, and the inherent \textit{sparsity} of failure-related signals in predominantly healthy systems.
Moreover, observational data often suffers from \textit{coverage gaps}, where certain system components, network segments, or application layers remain unmonitored due to instrumentation limitations, cost constraints, or architectural blind spots.
In practice, RCA systems frequently must operate with only \textit{partial observability}, where critical telemetry types may be missing entirely for specific services or time windows, forcing practitioners to infer root causes from incomplete evidence.
Furthermore, the sheer \textit{volume} and \textit{velocity} of telemetry generation pose significant challenges for real-time processing and storage.
Finally, the highly \textit{dynamic} nature of modern distributed systems means that both the system topology and telemetry patterns continuously evolve, complicating model stability and generalization.
These properties collectively force researchers to narrow their analytical scope to manageable subsets of the input space.

\paragraph{The Inherent Trade-off}
These data complexities give rise to a fundamental trade-off that governs all practical RCA systems.
Achieving desired \textbf{RCA effectiveness} is inextricably linked to the quantity and quality of available \textbf{observational data}.
However, increasing data granularity and collection scope to improve analytical power inevitably leads to higher \textbf{resource costs} for data ingestion, storage, transmission, and computation.
This creates a three-way tension that we term the \textbf{Effectiveness-Data-Cost Triangle}.
This tension forces researchers and practitioners to navigate competing priorities.
Some prioritize effectiveness by \textbf{expanding observational data} (\eg using eBPF), accepting higher costs.
Others focus on \textbf{enhancing data utilization} through advanced algorithms to extract more value from existing data.
A third approach emphasizes \textbf{improving computational efficiency} to enable more powerful analysis within the same cost envelope.
This fundamental tension explains why the idealized goal of RCA is often simplified into more tractable sub-problems.

\subsection{Seven Fundamental Goals of RCA} \label{sec:goals_definition}

The Effectiveness-Data-Cost triangle dictates that no single solution can simultaneously perfect all aspects of RCA.
Instead, existing research makes strategic compromises, focusing on tackling specific facets of this complex problem space.
This provides the central insight for our survey: the seemingly ad-hoc nature of RCA research is, in fact, a collection of rational, focused efforts to push the boundaries of this trade-off triangle along specific vectors.
These vectors can be distilled into seven fundamental goals that an ideal RCA system must pursue.
As introduced in Section~\ref{sec:introduction}, these goals directly map to the needs of different stages in the incident management lifecycle, where the ultimate objective is to reduce Mean Time to Recovery (MTTR) and extend Mean Time Between Failures (MTBF)~\cite{google_sre_incident_management}.
To provide a clear foundation for our taxonomy, we now define the scope and boundaries of each goal.

\begin{itemize}
    \item \textbf{Multi-dimensional Data Correlation.} This goal addresses the challenge of fusing heterogeneous telemetry data (\eg metrics, logs, traces) into a unified representation for analysis. Research in this area focuses on developing techniques for semantic alignment, such as creating shared embedding spaces or modeling cross-modal dependencies. It is distinct from \textit{robustness}, as it assumes data availability and prioritizes semantic integration over handling data imperfections.
    \item \textbf{Robustness.} This goal concerns the ability of an RCA model to function effectively with imperfect data, including noise, sparsity, and incompleteness. Methods in this category aim to infer causality from sparse signals, reconstruct incomplete system topologies, or denoise telemetry streams. It differs from \textit{adaptive learning} by addressing static data deficiencies, whereas the latter focuses on dynamic model adaptation to system evolution.
    \item \textbf{Adaptive Learning.} This goal focuses on enabling RCA models to evolve continuously in response to changes in the system's architecture, workload, or failure modes. The primary concern is developing mechanisms for online or incremental learning that obviate the need for complete model retraining, thus ensuring sustained performance in dynamic environments.
    \item \textbf{Real-time Performance.} As a non-functional requirement, this goal prioritizes the computational efficiency of RCA to ensure timely analysis during live incidents. The focus is on latency reduction through algorithmic optimization, parallelization, or approximation techniques. A method's contribution is measured by its speedup, distinguishing it from goals like \textit{interpretability}, which concerns the causal accuracy of the output rather than the speed of its generation.
    \item \textbf{Interpretability.} This goal aims to make RCA results understandable and trustworthy for human operators. The emphasis is on generating causally sound and logically coherent explanations, such as incident propagation graphs or natural language summaries. It seeks to answer the "why" and "how" of a failure, in contrast to \textit{multi-granularity}, which focuses on pinpointing the "where" at different abstraction levels.
    \item \textbf{Multi-granularity.} This goal is to achieve precise fault localization across multiple levels of abstraction, from high-level service dependencies down to specific code lines or configuration parameters. Its defining characteristic is the hierarchical depth and precision of the output, enabling drill-down analysis. This distinguishes it from the data-fusion focus of \textit{multi-dimensional data correlation} and the explanatory nature of \textit{interpretability}.
    \item \textbf{Actionability.} This goal focuses on translating diagnostic findings into concrete remedial actions. It bridges the gap between identifying a root cause and recommending a solution, such as suggesting a code rollback, generating a configuration patch, or retrieving relevant mitigation procedures from historical data. Unlike other goals centered on diagnosis, actionability is uniquely concerned with automated remediation.
\end{itemize}

\subsection{A Formal View of the RCA Problem} \label{sec:formal_rca}

\begin{figure}[t]
    \centering
    \includegraphics[width=1\linewidth]{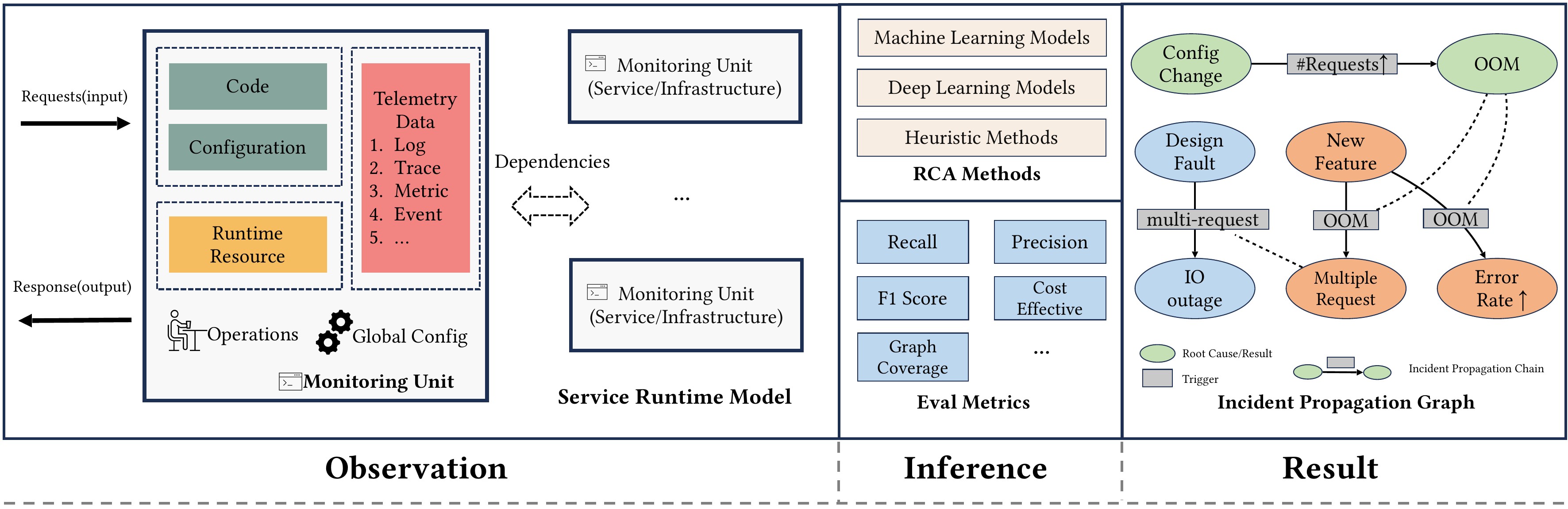}
    \caption{Overview of Root Cause Analysis. The \textbf{observation} space includes monitoring units as its basic building blocks. Each unit contains static files (such as configuration and code) and the runtime resources (for example, CPU, memory, and network) needed to transform these static programs into processes that interact with other services. Additionally, there is telemetry data that describes the runtime behavior of these processes, as well as the operations and global configurations maintained during maintenance. The \textbf{inference} component comprises various implemented methods alongside their corresponding evaluation metrics; its practical application is critically constrained by \textbf{efficiency} requirements (\eg real-time performance). Finally, the \textbf{output} identifies the incident's root causes and illustrates how the root cause propagates to the observed symptoms. This includes an incident propagation graph composed of an incident propagation chain. Each chain contains the root cause, trigger (optional), and result nodes.}
    \label{fig:overview}
\end{figure}

We formalize Root Cause Analysis (RCA) as a function $\mathcal{F}: \mathcal{O} \rightarrow \mathcal{G}$, where $\mathcal{O}$ denotes the hierarchical observation data and $\mathcal{G}$ represents the incident propagation graph.
As shown in Fig.~\ref{fig:overview}, the RCA process consists of three main components: observation (input), inference, and output.
This section first defines the input space $\mathcal{O}$ and the output space $\mathcal{G}$, and then discusses the role of this formalization as a conceptual framework for our survey.

\subsubsection{Input Space: Observation Data}

The input space $\mathcal{O}$ of RCA consists of the comprehensive telemetry data collected from various sources within a system.
We define the observation space as:

\begin{definition}[Input Space]
\[
\mathcal{O} = \{ \mathcal{L}, \mathcal{M}, \mathcal{T}, \mathcal{E}, \mathcal{D} \}
\]
\end{definition}

Where the components are defined as follows:
\begin{itemize}
    \item \(\mathcal{L}\) (\textbf{Logs}): Timestamped records of discrete events, crucial for debugging specific errors and capturing system behavior.
    \item \(\mathcal{M}\) (\textbf{Metrics}): Numerical measurements aggregated over time, offering high-level views of system health and performance indicators.
    \item \(\mathcal{T}\) (\textbf{Traces}): Data representing the end-to-end journey of requests across multiple services, providing context for distributed transactions through spans.
    \item \(\mathcal{E}\) (\textbf{Events}): Named occurrences at specific instants in time, representing discrete actions or state changes within a system~\cite{opentelemetry_semconv_events}. Events are the fundamental building blocks of telemetry and can be aggregated into logs, traces, and metrics~\cite{opentelemetry_logs_events,opentelemetry_trace_add_events,opentelemetry_metrics_event_model}.
    \item \(\mathcal{D}\) (\textbf{Supplementary Data}): Additional contextual information for RCA, including code, configuration files, and design documentation.
\end{itemize}

Together, these complementary data types enable operators to gain insights into system behavior, performance, and reliability, serving different aspects of the incident management lifecycle discussed in Section~\ref{sec:background}. Note that the observation space is often incomplete and noisy due to the inherent challenges in data collection and system complexity, as discussed in Section~\ref{sec:challenges}.

\subsubsection{Output Space: Incident Propagation Graph}

The output of RCA is an \textbf{Incident Propagation Graph}, which models the causal sequence of events that constitute an incident.
It is a composite of single or multiple failures, illustrating the relationships among events.
This graphical representation formalizes the mental model that Site Reliability Engineers (SREs) intuitively build when diagnosing cascading failures.

\begin{definition}[Output Space]
An Incident Propagation Graph is a directed acyclic graph (DAG) \(\mathcal{G} = ( \mathcal{V}, \mathcal{E})\), where $\mathcal{V}$ is a set of event nodes and $\mathcal{E} \subseteq \mathcal{V} \times \mathcal{V}$ represents the causal dependencies between them. Each node $v \in \mathcal{V}$ can be categorized by its role in the incident. We identify three primary roles: \textit{Root Cause, Trigger, and Symptom}.
To formalize the causal dynamics, we draw an analogy to chemical reactions.
The \textbf{root cause} ($r$) is a reactive component creating a latent failure condition (\eg a memory leak in code).
The optional \textbf{trigger} ($t$) acts as a catalyst that activates this condition (\eg a sudden traffic spike).
The \textbf{symptom} ($s$) is the observable manifestation of the failure (\eg an OOM error).
This formulation captures the essential temporal and causal relationship: the root cause establishes the precondition for failure, the trigger (if present) catalyzes the transition from latent to manifest failure, and the symptom is the observable outcome.
Similar to how catalysts are not consumed in a reaction, triggers are not root causes but facilitate the manifestation of symptoms, explaining why mitigating a trigger offers a quick fix while only resolving the root cause provides a permanent solution.
The optional nature of triggers parallels reactions that can proceed without a catalyst under specific conditions.
\end{definition}

\begin{figure}[t]
    \centering
    \includegraphics[width=0.9\textwidth]{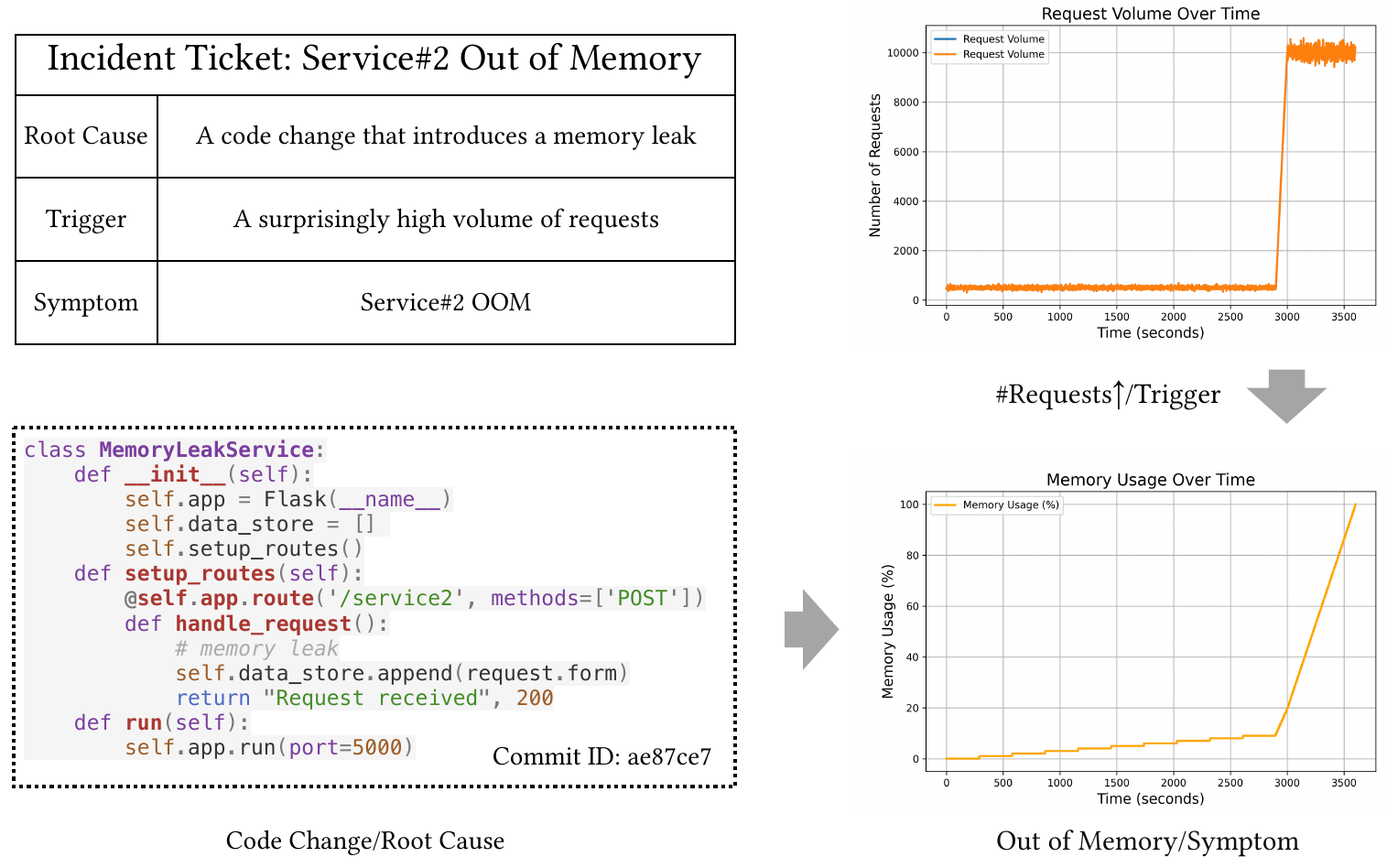}
    \caption{An Out-of-Memory (OOM) incident propagation chain. The OOM event in Service\#2 is the observed \textbf{symptom}. The \textbf{root cause} is the code change within Service\#2. The \textbf{trigger} is the increased traffic volume, which activates the latent flaw. Mitigating the trigger (\eg traffic reduction) provides a temporary fix, while resolving the root cause (\eg fixing the code) offers a permanent solution.}
    \label{fig:incident_propagation_chain}
\end{figure}

\begin{definition}[Root Cause Event Node]
\label{def:root_cause}
An event node \(r \in \mathcal{V}\) is a \emph{root cause} if it represents an initial fault or change and has an in-degree of zero in the incident propagation graph $\mathcal{G}$. That is, $\forall v \in \mathcal{V}, (v, r) \notin \mathcal{E}$. An incident may have one or more root causes. As illustrated in Fig.~\ref{fig:incident_propagation_chain}, the root cause is the fundamental flaw, like a memory leak in a recent code change. In practice, RCA systems often output a ranked list of \emph{root cause candidates (RCCs)}.
\end{definition}

\begin{definition}[Trigger Event Node]
\label{def:trigger}
An event node \(t \in \mathcal{V}\) is a \emph{trigger} if it is an event that activates a latent failure condition introduced by a root cause. A trigger is not a root cause itself but acts as a catalyst. The presence of a root cause alone may not lead to a failure; the failure manifests only when the trigger occurs. For example, a sudden traffic spike (the trigger) can cause a service with a memory leak (the root cause) to finally crash. Suppressing the trigger can quickly restore service, but does not fix the underlying issue. In many scenarios, a trigger may not be present or identifiable, making this component optional.
\end{definition}

\begin{definition}[Symptom Event Node]
\label{def:symptom}
An event node \(s \in \mathcal{V}\) is a \emph{symptom} if it represents an observed anomaly that directly initiates an incident response (\eg an OOM error or a service outage). It is the manifestation of a failure chain and, in many cases, corresponds to a node with an out-degree of zero in the graph $\mathcal{G}$.
\end{definition}

\begin{definition}[Intermediate Event Node]
An event node $v \in \mathcal{V}$ is an \emph{intermediate event} if it is neither a root cause nor a symptom that initiated the response. It acts as a symptom for its upstream events and a cause for its downstream events within a propagation chain.
\end{definition}

\subsubsection{Framework as a Unified Conceptual Model}

The formalization $\mathcal{F}: \mathcal{O} \rightarrow \mathcal{G}$ represents an \textbf{idealized and comprehensive goal} for RCA that encompasses the full spectrum of input-output paradigms found across existing literature.
For an SRE, identifying the root cause(s) answers the "what" question.
The incident propagation graph $\mathcal{G}$, particularly the propagation paths, answers the critical "how" and "why" questions.
The propagation path serves as the essential evidence chain to verify the correctness of the identified root cause, understand the incident's blast radius, and fundamentally prevent future recurrences.

However, reconstructing the complete propagation graph $\mathcal{G}$ faces enormous challenges, as detailed in Section~\ref{sec:challenges}.
Consequently, the vast majority of existing research simplifies the problem.
Most studies focus only on identifying a subset of event nodes from the graph, typically the root cause event node ($r$) and sometimes the symptom event node ($s$), rather than the full set of causal edges ($\mathcal{E}$) and trigger event nodes ($t$).
Their outputs often present a root cause at varying granularities (\eg service level, metric level, component level, or pod level) without the explanatory power of a full propagation path.
Notably, \textit{no existing work fully conforms to our idealized definition by outputting complete incident propagation paths}, highlighting the forward-looking nature of our survey.

While this formalization represents an idealized goal, its primary utility in this survey is to serve as a unified conceptual framework.
It allows us to deconstruct the diverse and seemingly disconnected problem formulations in existing literature.
By mapping each study's inputs and outputs onto our defined space ($\mathcal{O}$ and $\mathcal{G}$), we can precisely articulate \emph{how} they simplify the problem. For instance, many approaches focus solely on identifying the root-cause event node ($r$) rather than the entire graph.
This approach not only enables a more insightful comparison of different methods but also systematically charts the path forward for future, more comprehensive RCA research that moves from ''finding points'' to ''constructing graphs.''

This distinction is critical and forms a central theme of our analysis.
Mitigating the trigger, which is often the focus of SREs for immediate service restoration, offers a temporary fix to minimize MTTR.
In contrast, resolving the root cause, which is the primary goal for developers, provides a permanent solution to improve MTBF.
Recognizing this dichotomy is essential for understanding why different RCA methods produce outputs of varying nature and granularity, as they are implicitly optimized for different objectives within the incident management lifecycle.
\section{SURVEY METHODOLOGY}
\label{sec:surveymethod}

\subsection{SURVEY SCOPE}

Root cause analysis is a broad topic, applicable in a wide range of scenarios where it is essential to determine the causes behind a particular situation and how that situation arises. 
Examples include questions such as “Why an individual may have a high income”\cite{han2023root}, “Why intermittent slow queries occur in databases”\cite{ma2020diagnosing}, and “Why failures happen in microservices”~\cite{zheng2024mulan,wang2021groot}.

In this survey, we focus specifically on research that investigates the identification of root causes and how these causes contribute to observed behaviors in microservice systems, typically in the form of violations of expectations such as Service Level Objectives (SLOs). 
These systems are often characterized by complex environments with intricate service interactions. 
We exclude papers focused on fault localization~\cite{wong2016survey}, as this represents a narrower scope within the broader context of root cause analysis for microservices, primarily concerned with identifying vulnerable code segments. Notably, relevant fault injection benchmarks include Defects4j~\cite{just2014defects4j}.

\subsection{PAPER COLLECTION}

We conducted a systematic literature review of RCA research across four primary research domains.
To ensure comprehensive coverage, we systematically searched leading venues in Software Engineering, Systems and Distributed Computing, Artificial Intelligence and Databases, and other related disciplines.
The specific conferences and journals included in our search are presented in Table~\ref{tab:paper_collection_venues}.

\begin{table*}[t]
  \centering
  \caption{Venues Included in the Systematic Literature Review. We systematically searched \totalrcapaper top-tier conferences and journals across four primary research domains, covering Software Engineering, Systems and Distributed Computing, Artificial Intelligence and Databases, and related disciplines.}
  \label{tab:paper_collection_venues}
  \resizebox{\textwidth}{!}{%
  \begin{tabular}{l p{0.58\textwidth} p{0.28\textwidth}}
    \toprule
    \textbf{Domain} & \textbf{Conferences} & \textbf{Journals} \\
    \midrule
    Software Engineering & 
    ICSE, FSE/ESEC, ASE, ISSTA, ICST, ISSRE & 
    TOSEM, TSE, JSS \\
    \midrule
    Systems, Networking, \& Distributed Computing & 
    DSN, SIGCOMM, SIGMETRICS, EuroSys, ASPLOS, Middleware, ICDCS, INFOCOM, ICWS, CCGRID, GLOBECOM, ICC, IPCCC, IWQoS & 
    TSC, TDSC \\
    \midrule
    AI, Data Mining, \& Databases & 
    KDD, NeurIPS, ICLR, AAAI, SIGMOD, VLDB, CIKM, EMNLP, IJCNN & 
    --- \\
    \midrule
    Other Related Disciplines & 
    WWW, ICSOC, APNOMS, ICCBR, SMC, ISPA & 
    --- \\
    \bottomrule
  \end{tabular}%
  }
\end{table*}

To identify relevant literature, we initially conducted a manual search of the DBLP database, focusing on key conferences and using specific keywords such as "Root Cause", "Fault Localization," "Micro-Service", "Detection", and "Localization".
Recognizing the diverse nature of the RCA community, with its publications scattered across various venues and employing different terminologies, we aimed to ensure comprehensive coverage of the field.
To achieve this, we adopted a snowballing approach as recommended in \cite{jalali2012systematic}. This involved both backward and forward snowballing techniques:
\begin{itemize}
    \item \textbf{Backward Snowballing:} We reviewed the reference lists of each collected paper to identify additional relevant papers within our scope.
    \item \textbf{Forward Snowballing:} Using Google Scholar, we identified papers that cited our initially collected papers, thereby expanding our pool of relevant literature.
\end{itemize}

This iterative process was repeated until we reached a saturation point where no new relevant papers were identified. To maintain a high standard of quality, we ceased searching papers from non-top conferences and those with low citation counts from further reference list searches.

Furthermore, we limited our search to publications from the last ten years, as microservices have only gained significant traction after Google open-sourced Kubernetes~\cite{kubernetes}. Through this comprehensive search methodology, we identified and collected \totalrcapaper top papers directly related to Root Cause Analysis.

\subsection{PAPER ANALYSIS}

To ensure a thorough and rigorous analysis of the collected papers, we adopted a systematic approach closely tied to the theoretical framework proposed in Section~\ref{sec:preliminaries}. This process aimed to deconstruct and categorize each research work within our unified conceptual model of RCA.

The first two authors undertook an extensive reading and examination of the full text of each paper. For each paper, we extracted key information according to our idealized RCA definition ($\mathcal{F}: \mathcal{O} \rightarrow \mathcal{G}$). Specifically, we identified:
\begin{itemize}
    \item \textbf{Input Space ($\mathcal{O}$)}: What types of telemetry data does the research utilize (\eg logs, metrics, traces, events) and any supplementary data (\eg configurations, code changes).
    \item \textbf{Output Space (Simplified forms of $\mathcal{G}$)}: What does the research aim to identify? Root cause nodes ($r$), trigger nodes ($t$), symptom nodes ($s$), or partial paths between them? What is the granularity of the output (\eg service-level, instance-level, code-line-level)?
    \item \textbf{Core Methods and Challenges}: What methods or models does the research employ to address the problem? Which of our identified seven core goals does it primarily aim to tackle (\eg is it for improving \textbf{robustness} against noisy data, or achieving \textbf{real-time performance})?
    \item \textbf{Evaluation and Resources}: What evaluation methods and datasets does the research use? Are its code and datasets open-source?
\end{itemize}

Through this structured information extraction, we formalized each paper's specific problem formulation and mapped it to our goal-driven taxonomy. 
For example, a research work focusing on quickly localizing faulty services from noisy metric data would be categorized as primarily addressing \textbf{robustness} and \textbf{real-time performance} goals. 
This approach enables us to move beyond surface-level categorization based on input data types, revealing the deeper design philosophies and trade-offs underlying different research works.

When the two primary authors had differing interpretations or findings about paper categorization or information extraction, they conducted discussion sessions with additional co-authors. 
These discussions were instrumental in resolving disagreements and ensuring a consensus on the categorization of papers and the extracted data. 
The involvement of co-authors, who possess extensive expertise in RCA and microservices, helped maintain the accuracy and integrity of the analysis.

All authors independently reviewed the content to ensure the reliability and consistency of the survey's findings. 
This review process was designed to identify and correct any potential errors, inconsistencies, or omissions. 
By employing this rigorous multi-step analysis and review process, we ensured the credibility and robustness of our survey.
\section{Goal 1: Multi-dimensional Data Correlation}
\label{sec:multidimension}

As defined in the preliminaries (Section~\ref{sec:preliminaries}), the input space $\mathcal{O}$ for RCA comprises a heterogeneous collection of logs, metrics, traces, events, and supplementary data.
Each data type offers a unique yet siloed perspective on system behavior.
The goal of multi-dimensional data correlation is to overcome this fragmentation by fusing these diverse data sources into a unified and coherent view.
This fusion establishes a solid foundation for precise fault diagnosis.
This section focuses on the data fusion techniques themselves, examining how existing works ''translate'' and ''align'' these heterogeneous data silos.
We categorize these approaches into three mainstream paradigms: \textbf{fusion via unified representation learning}, \textbf{fusion via graph structures}, and the emerging \textbf{semantic fusion via large language models (LLMs)}.

\begin{table*}[t]
\centering
\caption{Overview of Data Fusion Paradigms and Representative Works for RCA.}
\label{tab:multidim_papers}
\footnotesize
\setlength{\tabcolsep}{4pt}
\renewcommand{\tabularxcolumn}[1]{m{#1}}
\begin{tabularx}{\textwidth}{@{}>{\centering\arraybackslash}m{0.19\textwidth}>{\centering\arraybackslash}m{0.16\textwidth}X@{}}
\toprule
\textbf{Fusion Paradigm} & \textbf{Method Category} & \textbf{Papers} \\
\midrule
Unified Representation & Eventization \& Embedding & DiagFusion~\cite{zhang2023robust}, Nezha~\cite{yu2023nezha}, UniDiag~\cite{zhang2024no}, DeepHunt~\cite{sun2025interpretable}, Chain-of-Event~\cite{yao2024chain} \\
\cmidrule(lr){2-3}
& Multi-modal Learning & \textbf{Contrastive:} MULAN~\cite{zheng2024mulan}, TVDiag~\cite{xie2025tvdiag} \\
\cmidrule(lr){3-3}
& & \textbf{Attention/Gating:} Eadro~\cite{lee2023eadro}, FAMOS~\cite{duan2025famos}, Medicine~\cite{tao2024giving} \\
\cmidrule(lr){3-3}
& & \textbf{Multi-stage:} ART~\cite{sun2024art} \\
\cmidrule(lr){3-3}
& & \textbf{Voting:} PDiagnose~\cite{hou2021Pdiagnose} \\
\midrule
Graph Structures & Heterogeneous Graph & TrinityRCL~\cite{gu2023trinityrcl}, CHASE~\cite{zhao2024chase}, FaaSRCA~\cite{huang2024faasrca}, SpanGraph~\cite{kong2024enhancing}, GIED~\cite{he2022graph} \\
\cmidrule(lr){2-3}
& Cross-layer Fusion & MicroRCA~\cite{wu2020microrca}, Microscope~\cite{lin2018microscope} \\
\cmidrule(lr){2-3}
& Graph Ensemble & FRL-MFPG~\cite{chen2023frl}, Chen et al.~\cite{chen2024graph}, CloudRCA~\cite{zhang2021cloudrca} \\
\cmidrule(lr){2-3}
& Domain-Specific & \textbf{Trace-based:} TraceAnomaly~\cite{liu2020unsupervised} \\
\cmidrule(lr){3-3}
& & \textbf{Data Structure:} TLCluster~\cite{sun2023trace}, LogKG~\cite{sui2023logkg} \\
\cmidrule(lr){3-3}
& & \textbf{Multi-metric:} AutoMAP~\cite{ma2020automap}, MS-Rank~\cite{ma2019ms}, CMMD~\cite{yan2022cmmd} \\
\cmidrule(lr){3-3}
& & \textbf{Temporal-structural:} GAMMA~\cite{somashekar2024gamma} \\
\cmidrule(lr){2-3}
& Case-based & MicroCBR~\cite{liu2022microcbr}, MicroTR~\cite{yao2025microtr}, SynthoDiag~\cite{zhang2024fault} \\
\cmidrule(lr){2-3}
& Correlation-based & HeMiRCA~\cite{zhu2024hemirca}, ICWS'20~\cite{wang2020root}, Log3C~\cite{he2018identifying}, MRCA~\cite{wang2024mrca} \\
\midrule
LLM-based Semantic & NL Transform & RCACopilot~\cite{chen2024automatic}, TrioXpert~\cite{sun2025trioxpert}, SCELM~\cite{sun2025multimodal}, X-lifecycle~\cite{goel2024x} \\
\cmidrule(lr){2-3}
& Knowledge-Data & \textbf{Documentation:} Atlas~\cite{xie2024cloud}, RealTCD~\cite{li2024realtcd} \\
\cmidrule(lr){3-3}
& & \textbf{Code:} Raccoon~\cite{zhao2023identifying}, COCA~\cite{li2025coca} \\
\cmidrule(lr){3-3}
& & \textbf{Multi-agent:} SynergyRCA~\cite{xiang2025simplifying}, KnowledgeMind~\cite{ren2025multi}, ThinkFL~\cite{zhang2025thinkfl} \\
\bottomrule
\end{tabularx}
\end{table*}

\subsection{Fusion via Unified Representation Learning}
\label{subsec:unified_representation}

This paradigm aims to map data from different modalities into a shared, low-dimensional vector space (embedding space), thereby enabling the analysis of originally incomparable data (such as discrete log events and continuous metric sequences) within a unified mathematical framework.

\subsubsection{Eventization and Embedding}

A core strategy involves abstracting all data into a unified "event" format and then leveraging this abstraction for further analysis.
For instance, \textit{DiagFusion}~\cite{zhang2023robust} and \textit{Nezha}~\cite{yu2023nezha} uniformly transform metric fluctuations, log entries, and trace anomalies into event sequences, employing FastText or similar models to generate embeddings for these events, thus enabling semantic similarity computation in vector space.
\textit{Chain-of-Event}~\cite{yao2024chain} also leverages this eventization framework to unify heterogeneous signals, but its primary goal is to use the resulting event stream to automatically learn a weighted, interpretable event-causal graph from historical data.
\textit{UniDiag}~\cite{zhang2024no} and \textit{DeepHunt}~\cite{sun2025interpretable} follow a similar approach but utilize the fused event sequences to construct temporal knowledge graphs or directly feed them into graph autoencoders to further learn normal system behavior patterns.

\subsubsection{Multi-modal Learning Architectures}

An alternative technical route employs \textbf{multi-modal learning architectures} that end-to-end learn fused representations through carefully designed neural networks.
These methods typically design specialized encoders for different data sources to preserve their unique characteristics, and then integrate information through a fusion module.
\textit{MULAN}~\cite{zheng2024mulan} and \textit{TVDiag}~\cite{xie2025tvdiag} introduce contrastive learning to not only learn modality-invariant features but also preserve modality-specific information, effectively preventing information loss during the fusion process.
\textit{Eadro}~\cite{lee2023eadro}, \textit{FAMOS}~\cite{duan2025famos}, and \textit{Medicine}~\cite{tao2024giving} leverage gating mechanisms or cross-attention to dynamically weight the contributions of different data sources, thereby generating more context-aware unified representations.
\textit{ART}~\cite{sun2024art} further employs a serialized multi-stage model (Transformer, GRU, GraphSAGE) to separately capture cross-channel, temporal, and topological dependencies, ultimately forming a "unified fault representation" that supports various downstream diagnostic tasks.
\textit{PDiagnose}~\cite{hou2021Pdiagnose} adopts a more lightweight approach, using a weighted voting mechanism to combine evidence from KPIs, traces, and logs.

\subsection{Fusion via Graph Structures}
\label{subsec:graph_fusion}

Unlike approaches that compress all information into vectors, graph-based fusion methods explicitly model different data entities and their relationships by constructing heterogeneous graphs.
In such graphs, nodes can represent services, instances, hosts, metric types, or even code entities, while edges denote call, hosting, causal, or correlation relationships among them.

\subsubsection{Graph Construction Strategies}

The key to these methods lies in graph construction strategies
\textit{TrinityRCL}~\cite{gu2023trinityrcl} and \textit{CHASE}~\cite{zhao2024chase} construct heterogeneous graphs containing multiple types of nodes, including services, hosts, metrics, and log anomalies, transforming the data fusion problem into a graph node feature learning problem.
\textit{FaaSRCA}~\cite{huang2024faasrca} extends this concept to serverless environments by building a "Global Call Graph" that integrates multi-modal data from both the application and the platform layers.
\textit{SpanGraph}~\cite{kong2024enhancing} focuses on a finer granularity, building a graph where nodes are spans and edges are invocations, with features enriched by metrics and configuration files.
\textit{GIED}~\cite{he2022graph} fuses numerical metrics and categorical service attributes into a unified graph representation to distinguish critical incidents from noise.
The contribution of \textit{MicroRCA}~\cite{wu2020microrca} and \textit{Microscope}~\cite{lin2018microscope} lies in their fusion of application-layer metrics with infrastructure-layer metrics, and their computation of correlations between cross-domain metrics to weight graph edges, thereby establishing cross-layer causal associations.
\textit{FRL-MFPG}~\cite{chen2023frl} fuses two distinct graphs (a call dependency graph extracted from traces and an association graph mined from historical faults) to construct a more comprehensive fault propagation model.
\textit{Chen et al.}~\cite{chen2024graph} achieves fusion by creating an ensemble model where different base learners specialize in either metric data or expert knowledge, with a meta-learner combining their outputs.
Similarly, \textit{CloudRCA}~\cite{zhang2021cloudpin} fuses KPIs, logs, and system topology into a Knowledge-informed Hierarchical Bayesian Network, using a probabilistic graphical model to perform inference.

\subsubsection{Domain-Specific Graph Integration}

Some works deeply integrate graph structures with specific data types.
For example, \textit{TraceAnomaly}~\cite{liu2020unsupervised} proposes Service Tracing Vectors (STV) that ingeniously fuse structured information (call paths) and numerical information (response times) into a single vector, which is then learned through Deep Bayesian Networks to capture patterns.
Others focus on creating integrated data structures.
\textit{TLCluster}~\cite{sun2023trace} creates a unified "trace log" data structure by combining traces with corresponding execution logs, enabling more precise instance-level fault localization through clustering.
\textit{LogKG}~\cite{sui2023logkg} focuses on integrating structured fields (such as component IDs) and unstructured content within logs by constructing a knowledge graph, thereby capturing a more comprehensive context than analyzing log messages alone.
\textit{AutoMAP}~\cite{ma2020automap} and \textit{MS-Rank}~\cite{ma2019ms} focus on fusing multiple performance metrics by constructing individual impact graphs for different metrics and then merging them into a composite graph to capture more complex performance issues.
\textit{CMMD}~\cite{yan2022cmmd} specifically addresses the fusion of different metrics by using a Graph Neural Network to automatically model the calculation relationships between fundamental metrics and derived KPIs.
\textit{GAMMA}~\cite{somashekar2024gamma} integrates temporal patterns from multi-variate metrics as node features into a structural dependency graph derived from traces.

\subsubsection{Case-based and Similarity-based Fusion}

Another distinct approach involves fusing heterogeneous data to enable case-based reasoning or diagnosis via similarity analysis.
These methods typically construct a knowledge graph or a "fault fingerprint" from multiple data sources for a given incident.
\textit{MicroCBR}~\cite{liu2022microcbr} combines anomalies from metrics, logs, and traces into such a fingerprint, which is then embedded into a spatio-temporal knowledge graph that also integrates system topology from a CMDB (Configuration Management Database).
This enables the retrieval of similar historical cases using a weighted longest common subsequence algorithm.
Similarly, \textit{MicroTR}~\cite{yao2025microtr} reproduces the execution state of transactions by building a multi-faceted knowledge graph that integrates temporal data, textual data from logs, call dependencies from traces, and performance metrics.
Diagnosis is performed via similarity analysis against a historical database of labeled transactions.
\textit{SynthoDiag}~\cite{zhang2024fault} also employs a knowledge graph to fuse execution logs, trace logs, and test case metadata for diagnosing test alarms.
It uses Knowledge Graph Embedding and Sentence-BERT to create unified vector representations, followed by a k-Nearest Neighbors classifier to determine the fault category.

\subsubsection{Correlation-based Fusion}

Beyond graph-based fusion, other works establish correlations between different data types to guide analysis.
\textit{HeMiRCA}~\cite{zhu2024hemirca} links a global anomaly score derived from traces to the behavior of individual metrics using Spearman correlation.
\textit{ICWS'20}~\cite{wang2020root} correlates log-derived anomaly scores with raw metric time series using Mutual Information.
\textit{Log3C}~\cite{he2018identifying} explicitly correlates the frequency of log clusters with KPI degradation to identify impactful problems.
\textit{MRCA}~\cite{wang2024mrca} first fuses logs and traces for anomaly detection, then uses metrics from anomalous services for causal analysis.

\subsection{Semantic Fusion via Large Language Models}
\label{subsec:llm_fusion}

The emergence of large language models (LLMs) has opened new pathways for multi-dimensional data correlation.
Their core advantage lies in powerful natural language understanding and generation capabilities, enabling them to act as "universal translators" that bridge the vast semantic gaps among unstructured, semi-structured, and structured data.

\subsubsection{Natural Language Transformation}

A mainstream approach involves \textbf{transforming all heterogeneous data into natural language descriptions} and providing them as context to LLMs for unified reasoning.
\textit{RCACopilot}~\cite{chen2024automatic}, \textit{TrioXpert}~\cite{sun2025trioxpert}, and \textit{SCELM}~\cite{sun2025multimodal} aggregate and summarize information from metrics, logs, traces, and change tickets into unified textual reports through automated workflows or dedicated data preprocessors, upon which LLMs perform root cause inference.
\textit{X-lifecycle}~\cite{goel2024x} further enriches this context by fusing incident data with historical knowledge bases and service dependency information.

\subsubsection{Knowledge-Data Integration}

A more sophisticated fusion approach leverages LLMs to \textbf{connect human knowledge with machine data}.
\textit{Atlas}~\cite{xie2024cloud} and \textit{RealTCD}~\cite{li2024realtcd} utilize LLMs to parse unstructured texts such as system design documents and operation manuals, extracting a priori causal knowledge and transforming it into structured graphs or constraints to guide and optimize traditional causal discovery algorithms based on numerical metrics, thereby achieving effective integration of human experience and machine data.
\textit{Raccoon}~\cite{zhao2023identifying} employs fault trees as an intermediate semantic layer, using LLMs to map user-reported natural language symptoms to fault tree nodes, which are then associated with specific code changes, successfully connecting the user perception layer with the engineering implementation layer.
\textit{SynergyRCA}~\cite{xiang2025simplifying} constructs runtime data into a knowledge graph, where LLMs actively explore and fuse information within the graph by generating query statements (Cypher), achieving dynamic, on-demand data fusion.
\textit{COCA}~\cite{li2025coca} fuses runtime issue reports with static source code by reconstructing execution paths, including across RPC boundaries, to create a code-aware context for the LLM.
\textit{KnowledgeMind}~\cite{ren2025multi} uses a multi-agent architecture where a verifier agent fuses processed information from specialized log, metric, and trace agents.
\textit{ThinkFL}~\cite{zhang2025thinkfl} achieves dynamic fusion through a reinforcement learning agent that adaptively decides whether to query trace or metric data at each step of its reasoning process.

\section{GOAL 2: ROBUSTNESS}\label{sec:goal2_robustness}

As established in Section~\ref{sec:preliminaries}, a fundamental challenge for any practical Root Cause Analysis (RCA) system is the imperfect nature of its input data $\mathcal{O}$.
Real-world telemetry is invariably marred by noise, incompleteness, or sparsity, which can severely undermine diagnostic accuracy.
The goal of \textbf{robustness}, therefore, is to ensure that an RCA system can maintain high accuracy and stability even when the quality of its input data is compromised.
To address this challenge, we organize the surveyed techniques into two primary strategies that tackle distinct facets of data imperfection.
The first, \textbf{handling observational blind spots}, focuses on reconstructing missing structural or behavioral information through inference, which is critical when dependency graphs are absent or key components lack direct monitoring.
The second, \textbf{ensuring algorithmic resilience}, concerns methods that are intrinsically designed to tolerate noise, data sparsity, and other inconsistencies within the available telemetry.
Together, these strategies enable RCA systems to deliver reliable insights under the adverse data conditions typical of production environments.

\begin{table*}[t]
    \centering
    \caption{Overview of Robustness Strategies and Representative Works for RCA.}
    \label{tab:robustness_summary}
    \footnotesize
    \setlength{\tabcolsep}{4pt}
    \renewcommand{\tabularxcolumn}[1]{m{#1}}
    \begin{tabularx}{\textwidth}{@{}>{\centering\arraybackslash}m{0.22\textwidth}>{\centering\arraybackslash}m{0.2\textwidth}X@{}}
        \toprule
        \textbf{Strategy} & \textbf{Sub-strategy} & \textbf{Papers} \\
        \midrule
        \multirow{8}{=}{\textbf{Handling Observational Blind Spots}} & \multirow{4}{=}{Structural Reconstruction} & \textbf{Constraint-based:} CloudRanger~\cite{wang2018cloudranger}, ServiceRank~\cite{ma2021servicerank}, MicroCause~\cite{meng2020localizing}, DyCause~\cite{pan2021faster}, GrayScope~\cite{zhang2024illuminating}, HRLHF~\cite{wang2023root}, ICWS'17~\cite{jia2017approach} \\
        \cmidrule(lr){3-3}
        & & \textbf{Learning-based:} CMDiagnostor~\cite{yu2023cmdiagnostor}, MonitorRank~\cite{kim2013root}, Murphy~\cite{harsh2023murphy}, CausalRCA~\cite{xin2023causalrca}, RUN~\cite{lin2024root} \\
        \cmidrule(lr){3-3}
        & & \textbf{Multi-modal/LLM:} MULAN~\cite{zheng2024mulan}, RealTCD~\cite{li2024realtcd} \\
        \cmidrule(lr){2-3}
        & Inference of Unobserved Components & LatentScope~\cite{xie2024microservice}, RCSF~\cite{wang2015methodology} \\
        \midrule
        \multirow{12}{=}{\textbf{Ensuring Algorithmic Resilience}} & \multirow{4}{=}{Tolerance to Sparsity \& Imbalance} & \textbf{Sparsity-specific:} MicroCU~\cite{jiang2023look}, SparseRCA~\cite{yao2024sparserca}, FaaSRCA~\cite{huang2024faasrca} \\
        \cmidrule(lr){3-3}
        & & \textbf{Imbalance (Learning):} DeepHunt~\cite{sun2025interpretable}, OCRCL~\cite{huang2024ocrcl}, TVdiag~\cite{xie2025tvdiag}, LasRCA~\cite{han2024potential}, SLIM~\cite{ren2024slim}, Raccoon~\cite{zhao2023identifying} \\
        \cmidrule(lr){3-3}
        & & \textbf{Imbalance (Data-level):} DiagFusion~\cite{zhang2023robust}, ICWS'20~\cite{wang2020root}, Medicine~\cite{tao2024giving} \\
        \cmidrule(lr){2-3}
        & \multirow{3}{=}{Statistical Aggregation \& Filtering} & \textbf{Robust Statistics:} CauseRank~\cite{lu2022generic}, MicroRank~\cite{yu2021microrank}, BARO~\cite{pham2024baro}, Squeeze~\cite{li2019generic}, ShapleyIQ~\cite{li2023shapleyiq}, $\epsilon$-Diagnosis~\cite{shan2019diagnosis} \\
        \cmidrule(lr){3-3}
        & & \textbf{Signal Enhancement:} FChain~\cite{nguyen2013fchain}, SwissLog~\cite{li2022swisslog}, TraceRCA~\cite{li2021practical}, WinG~\cite{yang2022robust} \\
        \cmidrule(lr){2-3}
        & Resilience to Novelty \& Dynamics & CloudRCA~\cite{zhang2021cloudrca}, MicroIRC~\cite{zhu2024microirc}, UniDiag~\cite{zhang2024no} \\
        \cmidrule(lr){2-3}
        & \multirow{2}{=}{Resilience via Agent Frameworks} & \textbf{Knowledge-guided:} Flow-of-Action~\cite{pei2025flow}, KnowledgeMind~\cite{ren2025multi} \\
        \cmidrule(lr){3-3}
        & & \textbf{Consensus-based:} mABC~\cite{zhang2024mabc}, RCAgent~\cite{wang2024rcagent} \\
        \bottomrule
    \end{tabularx}
\end{table*}

\subsection{Handling Observational Blind Spots}\label{sec:blindspot_robustness}

Observational blind spots are a primary challenge to robust RCA, arising when telemetry data $\mathcal{O}$ is structurally or behaviorally incomplete
This occurs when system topology information is missing, leaving a collection of isolated metrics, or when direct monitoring of a component is infeasible, providing only indirect signals
Robust systems overcome these blind spots using causal discovery and inference to reconstruct the missing information.

\subsubsection{Structural Reconstruction from Observational Data}

In dynamic microservice environments, maintaining an accurate, up-to-date dependency graph is a significant challenge.
Many robust RCA methods therefore operate without a predefined topology, instead inferring causal structure directly from observational data.
These approaches can be broadly categorized into statistical and learning-based methods.

Constraint-based statistical methods, particularly those based on the Peter-Clark (PC) algorithm, are widely used to construct causal graphs by testing for conditional independence among time-series metrics
\textit{CloudRanger}~\cite{wang2018cloudranger} and \textit{ServiceRank}~\cite{ma2021servicerank} exemplify this approach by dynamically building an "impact graph" from performance metrics, which then guides a random walk to identify the root cause
Recognizing that the standard PC algorithm's i.i.d assumption is violated by time-series data, \textit{MicroCause}~\cite{meng2020localizing} introduces the PCTS algorithm to explicitly model propagation delays, yielding a more robust causal graph for intra-service analysis
Similarly, Granger causality, which tests if one time series can forecast another, is another popular statistical tool
\textit{DyCause}~\cite{pan2021faster} applies it in sliding windows to build dynamic causality maps, while \textit{GrayScope}~\cite{zhang2024illuminating} and \textit{HRLHF}~\cite{wang2023root} integrate it with expert knowledge to constrain the search space and improve accuracy in noisy OS-level environments.
Other statistical approaches tackle imperfect data sources, such as mining control flows from interleaved logs that lack transaction identifiers~\cite{jia2017approach}.

Learning-based methods offer more flexibility in modeling complex, non-linear, and even cyclic dependencies that are common in enterprise systems but problematic for traditional statistical tests.
For instance, \textit{CMDiagnostor}~\cite{yu2023cmdiagnostor} uses regression on traffic patterns to resolve ambiguities in call metric data, thereby reconstructing a more accurate call graph.
Similarly, \textit{MonitorRank}~\cite{kim2013root} enhances robustness against imperfect static call graphs by using a probabilistic random walk and correcting for un-modeled correlations discovered in historical data.
\textit{Murphy}~\cite{harsh2023murphy} pioneers the use of Markov Random Fields (MRFs) to explicitly model and reason about cyclic dependencies, a critical capability for diagnosing issues like resource contention.
\textit{MULAN}~\cite{zheng2024mulan} enhances robustness against noisy data by using a KPI-aware attention mechanism to dynamically assess the reliability of different data modalities (metrics and logs) before fusing them into a unified causal graph, effectively down-weighting imperfect data sources.
\textit{RealTCD}~\cite{li2024realtcd} improves robustness by leveraging large language models to incorporate domain knowledge as a high-quality prior, guiding the causal discovery process and making it more resilient to imperfect interventional data where the interventional targets are unknown.
Other approaches leverage advanced neural architectures.
\textit{CausalRCA}~\cite{xin2023causalrca} employs a gradient-based method (DAG-GNN) to learn non-linear causal relationships, while \textit{RUN}~\cite{lin2024root} combines neural Granger causality with a novel contrastive learning scheme to robustly handle the complex periodicities in real-world metric data.

\subsubsection{Inference of Unobserved Components and States}

Beyond missing structure, RCA systems must often contend with unobserved components, where the true root cause candidate (RCC) lacks direct monitoring.
Robustness in this context means inferring the state of these hidden variables from their observable effects.
\textit{LatentScope}~\cite{xie2024microservice} directly tackles this problem by modeling unobservable RCCs as latent variables in a dual-space graph.
It then uses a regression-based algorithm to recognize interventions in this latent space, effectively identifying the unmonitored root cause from the behavior of related, observable metrics.
Similarly, \textit{RCSF}~\cite{wang2015methodology} infers the health of unmonitored functional components by analyzing their performance logs, enabling diagnosis in systems with incomplete monitoring coverage.

\subsection{Algorithmic Resilience to Data Imperfections}\label{sec:algorithmic_resilience}

Even when structural information is available, the data itself can be sparse, noisy, or imbalanced.
Algorithmic resilience refers to the intrinsic ability of a method to function effectively despite these data quality issues.

\subsubsection{Tolerance to Data Sparsity and Imbalance}

Data sparsity is a common problem, especially in testing environments or systems with low sampling rates.
Several methods are explicitly designed to operate under such constraints.
\textit{MicroCU}~\cite{jiang2023look} addresses extreme data sparsity (\eg 60-80\% missing) not by perfecting imputation, but by using "causal unimodalization" to calibrate and extract a reliable signal from the noisy causal curves derived from sparse data.
\textit{SparseRCA}~\cite{yao2024sparserca} targets sparse testing traces by performing analysis at the individual span level, avoiding the need for data aggregation altogether.
This challenge is particularly acute in serverless environments, where telemetry is discontinuous; \textit{FaaSRCA}~\cite{huang2024faasrca} addresses this by analyzing system snapshots rather than continuous time-series, making it robust to such transient data.

Data imbalance, particularly the scarcity of labeled failure instances, poses another major challenge.
\textit{DeepHunt}~\cite{sun2025interpretable} addresses the lack of labeled data through a self-supervised graph autoencoder, enabling a "zero-label cold start", and employs a data augmentation module that randomly masks input features to improve generalization from insufficient historical data.
Contrastive learning has emerged as a powerful technique to address this.
\textit{OCRCL}~\cite{huang2024ocrcl} and \textit{TVdiag}~\cite{xie2025tvdiag} both employ contrastive learning frameworks to learn discriminative representations from limited labeled data, effectively augmenting the sparse failure signals.
\textit{LasRCA}~\cite{han2024potential} takes this to an extreme in a one-shot learning scenario, using an LLM as a programmatic labeler to augment the training set for a small classifier.
Other methods tackle imbalance at the algorithmic level; for instance, \textit{SLIM}~\cite{ren2024slim} uses a submodular optimization framework to directly optimize the F1-score, making it inherently robust to the minority fault class.
Data augmentation is another common strategy, used by \textit{DiagFusion}~\cite{zhang2023robust} and \textit{ICWS'20}~\cite{wang2020root} to generate synthetic samples for rare failure types, preventing model bias.
In cases with very few historical incidents, \textit{Raccoon}~\cite{zhao2023identifying} generalizes from sparse 'seed' causal knowledge using a Tree GNN, enabling it to identify root causes even when direct historical evidence is unavailable.
\textit{Medicine}~\cite{tao2024giving} is explicitly designed to be robust against missing or low-quality data modalities through a parallel stream architecture and Multimodal Adaptive Optimization (MAO) module, allowing it to maintain high diagnostic accuracy even when one modality is compromised.

\subsubsection{Robustness Through Statistical Aggregation and Filtering}

Instead of relying on individual data points, which may be noisy, some methods achieve robustness by focusing on aggregate statistical patterns.
\textit{CauseRank}~\cite{lu2022generic} combats noise from numerous, similar abnormal metrics by grouping them before performing causal discovery, preventing the analysis from being misled by spurious correlations.
\textit{MicroRank}~\cite{yu2021microrank} enhances spectrum-based fault localization by using Personalized PageRank to weight traces based on their rarity, ensuring that unique, informative traces are not drowned out by common, noisy ones.
\textit{BARO}~\cite{pham2024baro} achieves robustness against inaccurate anomaly detection start times by using non-parametric statistics (median and IQR) instead of mean and standard deviation, making its analysis less sensitive to outliers.
\textit{Squeeze}~\cite{li2019generic} is designed to be robust to anomalies of both significant and insignificant magnitudes by using a Generalized Potential Score (GPS) that is less sensitive to cumulative forecast errors.
\textit{ShapleyIQ}~\cite{li2023shapleyiq} provides robustness in scenarios with multiple coexisting root causes, as its Shapley value framework inherently considers all combinations of factors, allowing it to accurately quantify the influence of concurrent faults
Similarly, $\epsilon$-\textit{Diagnosis}~\cite{shan2019diagnosis} uses distribution-free e-statistics to compare time-series distributions, making it well-suited for the heavy-tailed and high-variance data characteristic of long-tail latency events.

Other methods use filtering techniques to improve the signal-to-noise ratio.
\textit{FChain}~\cite{nguyen2013fchain} dynamically adjusts anomaly detection thresholds based on a metric's predictability (via FFT), preventing normal workload fluctuations from being mistaken for faults.
Others filter noise at the data source; \textit{SwissLog}~\cite{li2022swisslog} achieves robustness to changing log formats through a dictionary-based parser and semantic analysis, while \textit{TraceRCA}~\cite{li2021practical} employs adaptive feature selection to dynamically ignore irrelevant metrics during an incident.
\textit{WinG}~\cite{yang2022robust} uses Dynamic Time Warping (DTW) to compare temporal sequences, providing a non-linear alignment that is resilient to normal variations in service invocation patterns.

\subsubsection{Resilience to Novelty and System Dynamics}

A critical aspect of robustness is the ability to diagnose novel faults not seen during training and to adapt to dynamic system topologies.
Several methods build this resilience into their model architecture.
\textit{CloudRCA}~\cite{zhang2021cloudrca} uses a hierarchical Bayesian network that can generalize to identify the correct fault module even if the specific fault type is new.
\textit{MicroIRC}~\cite{zhu2024microirc} combines a supervised GNN with an unsupervised random walk on a real-time graph, reducing dependency on a static set of failure signatures and making it resilient to both new anomaly types and dynamic instance scaling.
Similarly, \textit{UniDiag}~\cite{zhang2024no} is designed to diagnose previously unseen failure classes and can maintain functionality even when certain data modalities are missing, demonstrating resilience to both novelty and incomplete data.

\subsubsection{Resilience via Multi-Agent and LLM-based Frameworks}

The recent adoption of Large Language Models (LLMs) in RCA has introduced a new source of imperfection: model hallucination and instability.
Robust frameworks in this domain focus on constraining and validating the LLM's reasoning process.
\textit{Flow-of-Action}~\cite{pei2025flow} and \textit{KnowledgeMind}~\cite{ren2025multi} use expert knowledge, encoded as Standard Operating Procedures (SOPs) or rules, to guide the agent's exploration and provide a reward mechanism, effectively grounding the model and preventing it from pursuing irrelevant diagnostic paths.
A complementary approach is to use consensus.
\textit{mABC}~\cite{zhang2024mabc} employs a blockchain-inspired voting mechanism where multiple specialized agents must reach a consensus, preventing a single agent's error from derailing the entire analysis.
Similarly, \textit{RCAgent}~\cite{wang2024rcagent} uses Trajectory-level Self-Consistency (TSC) to aggregate multiple reasoning paths into a more reliable final answer, enhancing robustness even when using less powerful, locally-hosted LLMs.

\section{GOAL 3: ADAPTIVE LEARNING}\label{sec:goal3_adaptive}

As established in Section~\ref{sec:preliminaries}, the highly dynamic nature of modern systems complicates model stability and generalization, demanding that RCA systems evolve in tandem.
This section addresses this challenge through our third goal, \textbf{Adaptive Learning}, which enables RCA systems to \textbf{continuously adapt to dynamic conditions, novel failures, and evolving system topologies without costly retraining}.
Unlike robustness (Goal 2), which handles static data imperfections, adaptive learning focuses on the model's dynamic response to temporal system changes, such as service deployments and workload shifts.
We survey three primary adaptation paradigms: \textbf{Incremental Model Evolution}, for updating models with new data streams; \textbf{Rapid Generalization}, for adapting to new environments with minimal data; and \textbf{Intelligent Policy and Knowledge Adaptation}, for dynamically adjusting analytical strategies.

\begin{table*}[t]
    \centering
    \caption{Overview of Adaptive Learning Paradigms and Representative Works for RCA.}
    \footnotesize
    \setlength{\tabcolsep}{4pt}
    \renewcommand{\tabularxcolumn}[1]{m{#1}}
    \begin{tabularx}{\textwidth}{@{}>{\centering\arraybackslash}m{0.22\textwidth}>{\centering\arraybackslash}m{0.25\textwidth}X@{}}
        \toprule
        \textbf{Paradigm} & \textbf{Strategy} & \textbf{Papers} \\
        \midrule
        \multirow{4}{=}{\textbf{Incremental Model Evolution}} & Graph-based Incremental Updates & CORAL~\cite{wang2023incremental}, Sage~\cite{gan2021sage}, DGERCL~\cite{cheng2024dgercl} \\
        \cmidrule(lr){2-3}
        & Feedback-driven Refinement & DeepHunt~\cite{sun2025interpretable}, IPCCC'16~\cite{nie2016mining}, ServerRCA~\cite{shi2023serverrca}, Chain-of-Event~\cite{yao2024chain}, TraceStream~\cite{zhou2023tracestream}, UniDiag~\cite{zhang2024no} \\
        \cmidrule(lr){2-3}
        & Self-optimization Mechanisms & AutoMAP~\cite{ma2020automap}, MS-Rank~\cite{ma2019ms}, Medicine~\cite{tao2024giving} \\
        \cmidrule(lr){2-3}
        & Online Learning with Memory Mechanisms & OCRCL~\cite{huang2024ocrcl}, CloudPD~\cite{sharma2013cloudpd}, MicroSketch~\cite{li2022microsketch} \\
        \midrule
        \textbf{Rapid Generalization} & Few-shot \& Zero-shot Learning & Sleuth~\cite{gan2023sleuth}, SpanGraph~\cite{kong2024enhancing}, SparseRCA~\cite{yao2024sparserca} \\
        \midrule
        \multirow{2}{=}{\textbf{Intelligent Policy \& Knowledge Adaptation}} & Reinforcement Learning for Policy Adaptation & TraceDiag~\cite{ding2023tracediag}, ThinkFL~\cite{zhang2025thinkfl}, HRLHF~\cite{wang2023root} \\
        \cmidrule(lr){2-3}
        & Knowledge Retrieval \& Dynamic Reasoning (RAG) & ICLRCA~\cite{zhang2024automated}, SCELM~\cite{sun2025multimodal}, Xpert~\cite{jiang2024xpert} \\
        \bottomrule
    \end{tabularx}
\end{table*}

\subsection{Incremental Model Evolution}
\label{subsec:incremental_evolution}

This paradigm focuses on updating existing models with new data streams, ensuring they remain current without discarding previously learned knowledge, a challenge often referred to as catastrophic forgetting. 
These approaches are crucial for monitoring the gradual shift in system behavior and architecture.

\subsubsection{Graph-based Incremental Updates}

A prominent strategy involves incrementally updating graph-based models to reflect evolving system dependencies. 
\textit{CORAL}~\cite{wang2023incremental} and \textit{Sage}~\cite{gan2021sage} both address architectural dynamism by updating graph representations. 
CORAL achieves this by disentangling the causal graph into state-invariant and state-dependent components, allowing for efficient updates on the dynamic part.
Its LSTM component captures long-term temporal patterns, while the VGAE locally updates graph structures as services are added, removed, or dependencies modified, significantly reducing computational overhead while maintaining accuracy.
Sage, in contrast, decomposes its GVAE model on a per-microservice basis following a Causal Bayesian Network structure, enabling selective partial retraining of only the components affected by a change.
This targeted approach reduces retraining time while preserving causal integrity.

Similarly, \textit{DGERCL}~\cite{cheng2024dgercl} employ online learning to process streams of multi-modal data, incrementally updating their graph structures and node embeddings to capture temporal dynamics.
DGERCL processes continuous streams of invocation events through an LSTM, dynamically updating microservice node embeddings and employing a self-attention mechanism to weigh the importance of different metrics during the evolution of system behavior.

\subsubsection{Feedback-driven Refinement}

Another class of methods achieves adaptation through feedback-driven refinement, incorporating human-in-the-loop or automated validation steps to continuously fine-tune the model. 
\textit{DeepHunt}~\cite{sun2025interpretable}, \textit{IPCCC'16}~\cite{nie2016mining}, and \textit{ServerRCA}~\cite{shi2023serverrca} exemplify this by using operator feedback to correct or confirm diagnoses, which then serves as new labeled data to refine model parameters or knowledge graphs. 
DeepHunt employs a feedback mechanism to fine-tune its root cause scorer via a ranking-oriented loss function, allowing continuous adaptation to new failures.
IPCCC'16 uses a closed-loop approach where operators label causal rules, which are then used to train a Random Forest classifier that updates rule weights in the causality graph.
ServerRCA incorporates a human-in-the-loop mechanism specifically for handling unseen fault events, flagging them for expert review and adding newly labeled events to its knowledge repository without full model retraining.
\textit{Chain-of-Event}~\cite{yao2024chain} also supports adaptation by enabling the model to be retrained as new labeled incident data accumulates over time, allowing its event-causal graph to evolve with the system.
\textit{TraceStream}~\cite{zhou2023tracestream} applies this concept at a cluster level, allowing operators to label groups of anomalous traces efficiently.
By utilizing an online data stream clustering algorithm (DenStream), the model continuously updates to handle concept drift caused by system updates, making operator feedback more scalable than per-trace labeling.
\textit{UniDiag}~\cite{zhang2024no} supports incremental learning by flagging new failure embeddings that are significantly distant from all known clusters, creating new failure type clusters that can be labeled by operators without requiring a full retraining cycle.

\subsubsection{Self-optimization Mechanisms}

A more automated variant is self-optimization, where models adjust their internal parameters based on diagnostic performance without explicit human intervention. 
\textit{AutoMAP}~\cite{ma2020automap} and the two variants of \textit{MS-Rank}~\cite{ma2019ms} dynamically update the weights assigned to different metrics based on their historical success in localizing faults.
AutoMAP maintains a history of incidents and calculates similarity to past incidents, using confirmed outcomes to update metric weights for current diagnoses.
MS-Rank employs a self-adaptive mechanism that evaluates the precision of each diagnosis result and updates the confidence weights of metrics accordingly, enabling the framework to optimize itself over multiple incidents.
\textit{Medicine}~\cite{tao2024giving} employs a sophisticated form of this through its Multimodal Adaptive Optimization (MAO) module, which dynamically balances the learning rates across different data modalities during training.
By evaluating each modality's contribution, it suppresses gradients for high-performing modalities and enhances features for underperforming ones, preventing a dominant data source from suppressing others and ensuring all modalities contribute effectively.

\subsubsection{Online Learning with Memory Mechanisms}

Some approaches utilize online learning with memory mechanisms to balance stability and plasticity. 
\textit{OCRCL}~\cite{huang2024ocrcl} uses a memory replay strategy to incrementally train its contrastive learning model on new business incidents without forgetting past knowledge, addressing the scarcity of historical data and enabling real-time model updates in evolving systems.
\textit{CloudPD}~\cite{sharma2013cloudpd} and \textit{MicroSketch}~\cite{li2022microsketch} maintain an adaptive model of normal behavior by continuously updating it with recent data streams.
CloudPD employs a k-Nearest Neighbors model on an operating context defined by host metrics, while MicroSketch uses a Robust Random Cut Forest (RRCForest) that processes data as a stream and dynamically adapts to changes such as service auto-scaling or updates without offline retraining.

\subsection{Rapid Generalization through Few-shot Learning}
\label{subsec:rapid_generalization}

Rapid generalization addresses the challenge of quickly adapting to entirely new environments or failure modes with minimal training data.
This capability is crucial for organizations deploying RCA systems across diverse microservice environments or when encountering novel failure patterns not seen in historical data.

\textit{Sleuth}~\cite{gan2023sleuth} represents a significant advancement over traditional incremental approaches by utilizing Graph Neural Networks (GNNs) to enable few-shot and zero-shot learning.
Unlike Sage~\cite{gan2021sage}, which relies on GVAE and CBN for incremental updates, Sleuth's GNN architecture captures generalizable patterns across different microservices, reducing the need for extensive retraining.
This approach makes Sleuth particularly effective in dynamic environments where rapid deployment and high accuracy are essential, as it can quickly adapt to new microservice applications with minimal data requirements.

\textit{SpanGraph}~\cite{kong2024enhancing} demonstrates exceptional performance in few-shot learning scenarios, achieving high F1 scores of 93\% and 88.95\% on SockShop~\cite{sockshop} and Trainticket~\cite{trainticket} datasets respectively with just 1\% of the training data.
The model's performance consistently improved across precision, recall, and F1-score as data proportion increased, highlighting its efficiency and reliability in generalizing with minimal data.
This capability makes SpanGraph particularly valuable for fault localization in microservices systems where comprehensive training data is scarce.

\textit{SparseRCA}~\cite{yao2024sparserca} is specifically designed for sparse data environments, particularly in testing scenarios, and demonstrates strong adaptability to knowledge obsolescence.
It addresses the challenge of frequent system upgrades that lead to constantly emerging new trace structures by estimating expected latency for entirely new patterns through parameter extrapolation from the most similar known structures.
This enables the model to perform accurate RCA on previously unencountered trace structures without requiring retraining.

\subsection{Intelligent Policy and Knowledge Adaptation}
\label{subsec:policy_adaptation}

This paradigm moves beyond updating data representations to adapting the analytical \textit{strategy} itself. 
This is achieved through Reinforcement Learning (RL) or by leveraging the dynamic reasoning and retrieval capabilities of Large Language Models (LLMs).

\subsubsection{Reinforcement Learning for Policy Adaptation}

RL-based policy adaptation enables systems to learn the optimal sequence of analytical actions.
\textit{TraceDiag}~\cite{ding2023tracediag} exemplifies this approach by using RL to automatically learn adaptive pruning policies for service dependency graphs.
The system employs Proximal Policy Optimization (PPO) to learn a policy represented as a filtering tree, which selectively eliminates redundant components based on latency, anomaly indicators, and correlation metrics.
This learned policy adapts to changing system characteristics, ensuring that only the most relevant components are retained for subsequent causal analysis.
\textit{ThinkFL}~\cite{zhang2025thinkfl} advances this paradigm further by using reinforcement fine-tuning to teach a lightweight LLM to autonomously discover optimal reasoning paths.
Rather than following a rigid workflow, ThinkFL's "Recursion-of-Thought" actor dynamically decides which data tools to query through a progressive Group Relative Policy Optimization (GRPO) training process.
A multi-factor reward function evaluates both the accuracy of the final root cause ranking and the quality of the reasoning path, guiding the LLM to learn interpretable and efficient localization strategies.
\textit{HRLHF}~\cite{wang2023root} integrates human feedback into the reinforcement learning process, drawing inspiration from RLHF techniques used to align large language models.
By incorporating expert guidance, HRLHF constructs dependency graphs with high accuracy while minimizing human intervention requirements, effectively learning generalizable patterns for autonomous operation while adapting its analytical approach based on domain expertise.

\subsubsection{Knowledge Retrieval and Dynamic Reasoning}

A more recent and powerful approach for adaptation is Retrieval-Augmented Generation (RAG) with LLMs.
This method avoids model retraining entirely by dynamically retrieving relevant, up-to-date information from an external knowledge base at inference time.

\textit{ICLRCA}~\cite{zhang2024automated}, \textit{SCELM}~\cite{sun2025multimodal}, and \textit{Xpert}~\cite{jiang2024xpert} all leverage this technique by maintaining continuously updated vector databases of historical incidents, their root causes, and associated queries or solutions.
When a new incident occurs, the most semantically similar past cases are retrieved and injected into the LLM's context, allowing the model to reason using the most current knowledge without any modification to its internal weights.
ICLRCA uses this approach with GPT-4 for automated root cause analysis, outperforming fine-tuned models while avoiding computational costs and data staleness issues.
SCELM extends this to unify erroneous change detection, failure triage, and root cause change analysis into a single automated pipeline, processing multimodal data and accessing operational knowledge dynamically.
Xpert applies the same paradigm to generate customized domain-specific queries (\eg KQL) for incident investigation, with the vector database continuously updated as new incident-query pairs are resolved.
This RAG-based approach makes the system inherently adaptive to new failure patterns as soon as they are documented in the knowledge base, representing a shift from model adaptation to knowledge adaptation.

\section{Goal 4: Real-time Performance}
\label{sec:goal4_realtime}

As defined in Section~\ref{sec:preliminaries}, real-time performance is a critical non-functional requirement for practical Root Cause Analysis (RCA) systems.
This goal directly addresses the operational imperative to minimize Mean Time to Recovery (MTTR) by ensuring diagnostic completion within seconds or minutes.
Achieving this objective requires navigating the fundamental trade-off between analytical depth and response time, a core tension encapsulated in the Effectiveness-Data-Cost Triangle (Section~\ref{sec:preliminaries}).
This challenge is exacerbated by the massive data volumes and architectural scale of modern cloud environments.
To address this, the literature has converged on three primary strategies, which we review in this section: 
\textbf{(1) Computational Optimization}, which reduces the problem's search space; 
\textbf{(2) Efficient Algorithmic Design}, which employs intrinsically fast algorithms; and 
\textbf{(3) Architectural Acceleration}, which uses system-level parallelism and incremental computation.

\begin{table*}[t]
    \centering
    \caption{Overview of Real-time Performance Paradigms and Representative Works for RCA.}
    \label{tab:realtime_summary}
    \footnotesize
    \setlength{\tabcolsep}{4pt}
    \renewcommand{\tabularxcolumn}[1]{m{#1}}
    \begin{tabularx}{\textwidth}{@{}>{\raggedright\arraybackslash}m{0.2\textwidth}>{\raggedright\arraybackslash}m{0.28\textwidth}X@{}}
        \toprule
        \textbf{Paradigm} & \textbf{Strategy} & \textbf{Papers} \\
        \midrule
        \multirow{3}{=}{\textbf{Computational Optimization}} & Heuristic \& Statistical Pruning & MicroHECL~\cite{liu2021microhecl}, TraceDiag~\cite{ding2023tracediag}, GIED~\cite{he2022graph}, TraceContrast~\cite{zhang2024trace}, PatternMatcher~\cite{wu2021identifying}, Onion~\cite{zhang2021onion}, COMET~\cite{wang2024large} \\
        \cmidrule(lr){2-3}
        & Dimensionality Reduction \& Attribute Selection & RCOAS~\cite{cheng2022effective}, TS-InvarNet~\cite{hu2022ts}, CMMD~\cite{yan2022cmmd}, Squeeze~\cite{li2019generic} \\
        \cmidrule(lr){2-3}
        & Hierarchical \& Localized Search & HALO~\cite{zhang2021halo}, RCD~\cite{ikram2022root}, CauseRank~\cite{lu2022generic}, DyCause~\cite{pan2021faster} \\
        \midrule
        \multirow{3}{=}{\textbf{Efficient Algorithmic Design}} & Efficient Data Summarization \& Representation & MicroSketch~\cite{li2022microsketch}, KPIRoot~\cite{gu2024kpiroot}, Log3C~\cite{he2018identifying} \\
        \cmidrule(lr){2-3}
        & Low-Complexity Analytical Algorithms & ShapleyIQ~\cite{li2023shapleyiq}, Minesweeper~\cite{murali2021scalable}, PDiagnose~\cite{hou2021Pdiagnose}, $\epsilon$-Diagnosis~\cite{shan2019diagnosis}, FluxRank~\cite{liu2019fluxrank} \\
        \cmidrule(lr){2-3}
        & Lightweight Model-Based Approaches & SLIM~\cite{ren2024slim}, MonitorRank~\cite{kim2013root} \\
        \midrule
        \multirow{5}{=}{\textbf{Architectural Acceleration}} & Parallel \& Distributed Processing & TraceContrast~\cite{zhang2024trace}, FacGraph~\cite{lin2018facgraph}, Microscope~\cite{lin2018microscope}, Sage~\cite{gan2021sage}, Murphy~\cite{harsh2023murphy}, CIRCA~\cite{li2022causal} \\
        \cmidrule(lr){2-3}
        & Incremental \& Online Learning & CORAL~\cite{wang2023incremental}, MRCA~\cite{wang2024mrca} \\
        \cmidrule(lr){2-3}
        & Multi-Stage \& Hierarchical Architectures & ChangeRCA~\cite{yu2024changerca}, CloudPD~\cite{sharma2013cloudpd}, FChain~\cite{nguyen2013fchain}, Roots~\cite{jayathilaka2017performance} \\
        \cmidrule(lr){2-3}
        & Efficient Integration Frameworks & UniDiag~\cite{zhang2024no}, Groot~\cite{wang2021groot}, MicroDig~\cite{tao2024diagnosing} \\
        \cmidrule(lr){2-3}
        & Specialized Approaches & ModelCoder~\cite{cai2021modelcoder}, TraceRank~\cite{yu2023tracerank}, TraceStream~\cite{zhou2023tracestream}, GLOBECOM’18~\cite{ikeuchi2018root} \\
        \midrule
        \multirow{2}{=}{\textbf{LLM-Enhanced Efficiency}} & Context Window \& Token Optimization & KnowledgeMind~\cite{ren2025multi}, OpenRCA~\cite{xu2025openrca}, XPERT~\cite{sun2025trioxpert} \\
        \cmidrule(lr){2-3}
        & Model Optimization \& Fine-Tuning & eARCO~\cite{goel2025earco}, ThinkFL~\cite{zhang2025thinkfl} \\
        \bottomrule
    \end{tabularx}
\end{table*}

\subsection{Computational Optimization via Search Space Reduction}
\label{subsec:search_space_reduction}

The most prevalent strategy for achieving real-time performance is to aggressively reduce the computational search space.
Recognizing that telemetry data is often sparse in failure-related signals, these methods employ sophisticated pruning and filtering techniques to focus analytical resources on the most relevant data subsets.

\subsubsection{Heuristic and Statistical Pruning}

A primary technique is to eliminate irrelevant data early in the pipeline based on statistical significance or correlation thresholds.
\textit{MicroHECL}~\cite{liu2021microhecl} prunes anomaly propagation chains by assessing the Pearson correlation of metrics between successive service calls, eliminating edges below certain thresholds.
\textit{TraceDiag}~\cite{ding2023tracediag} employs a reinforcement learning agent to learn an optimal pruning policy based on latency, anomaly, and correlation criteria, dramatically reducing the size of the dependency graph before causal analysis.
Similarly, \textit{GIED}~\cite{he2022graph} leverages DBSCAN clustering and influence topology filtering to eliminate nodes with low structural significance, while \textit{TraceContrast}~\cite{zhang2024trace} employs chi-square and minimum support pruning to filter out statistically insignificant patterns.

Several systems incorporate multi-stage filtering to progressively narrow the scope of the analysis.
\textit{PatternMatcher}~\cite{wu2021identifying} uses a lightweight KS-test for initial filtering before pattern classification, while \textit{Onion}~\cite{zhang2021onion} implements downward-closure based pruning during clustering to avoid generating trivial log groups.
\textit{COMET}~\cite{wang2024large} employs an AutoExtractor to filter and select the most relevant logs from various sources, creating a concise information subset for downstream analysis.

\subsubsection{Dimensionality Reduction and Attribute Selection}

Another group of methods focuses on reducing the dimensionality of the search space itself.
\textit{RCOAS}~\cite{cheng2022effective} introduces an attribute selection pre-processing step that filters irrelevant dimensions, enabling downstream multi-dimensional analysis algorithms to run orders of magnitude faster (\eg reducing \textit{HALO}'s execution time from 199.6s to 5.9s).
This approach combines rule-based filtering with an improved Logistic Iterative Relief (LIR) algorithm that is robust to data imbalance.
\textit{TS-InvarNet}~\cite{hu2022ts} uses shape-based clustering (HDBSCAN) on KPIs to reduce redundancy, significantly decreasing the number of pairwise invariants that need to be mined and monitored.

For multi-dimensional analysis, \textit{CMMD}~\cite{yan2022cmmd} and \textit{Squeeze}~\cite{li2019generic} tackle the combinatorial explosion by replacing exhaustive searches with more efficient alternatives.
\textit{CMMD} employs a genetic algorithm with an attention-based filtering mechanism to handle search spaces of up to $10^5$ dimension value combinations, while \textit{Squeeze} proposes a novel "bottom-up then top-down" search strategy that first clusters potential anomalies and then efficiently searches within clusters using the Generalized Potential Score (GPS) heuristic.

\subsubsection{Hierarchical and Localized Search}

A more structured approach to pruning involves leveraging hierarchical or causal structures.
\textit{HALO}~\cite{zhang2021halo} implements a sophisticated two-phase approach that leverages hierarchical attribute relationships, constructing an Attribute Hierarchy Graph (AHG) to organize attributes hierarchically.
The system performs an efficient attribute-level search phase followed by a value-level search with adaptive early-stopping mechanisms, enabling it to scale to 1.2 million records while maintaining real-time performance.

\textit{RCD}~\cite{ikram2022root} achieves significant speedup by avoiding the construction of a complete causal graph.
Its hierarchical, divide-and-conquer algorithm performs localized causal discovery on small subsets of metrics, applying the $\Psi$-PC algorithm to identify potential root causes within each subset and then recursively combining results.
This method dramatically reduces the number of required conditional independence tests, making the approach orders of magnitude faster than baseline causal discovery methods and practical for large-scale systems with thousands of metrics.
\textit{CauseRank}~\cite{lu2022generic} similarly reduces causal graph complexity by operating on metric groups rather than individual metrics, incorporating domain knowledge to improve both accuracy and efficiency.
\textit{DyCause}~\cite{pan2021faster} also accelerates its dynamic causality discovery through optimized pruning strategies and by confining its analysis to specific, detected anomaly intervals, making it significantly faster than baseline methods.

\subsection{Efficient Algorithmic Design}
\label{subsec:efficient_algorithms}

Beyond pruning, a second major line of work focuses on designing algorithms and data structures that are intrinsically lightweight and computationally efficient.
This approach prioritizes low-overhead computation from the ground up, enabling RCA systems to handle large-scale data without sacrificing speed.

\subsubsection{Efficient Data Summarization and Representation}

One prominent technique is the use of compact data representations that preserve essential diagnostic information while drastically reducing computational cost.
\textit{MicroSketch}~\cite{li2022microsketch} employs an extended DDSketch data structure to summarize trace latency distributions with sublinear space and linear time complexity, enabling it to analyze 10,000 traces in approximately 1.1 seconds, which is at least 60 times faster than competing methods.
Similarly, \textit{KPIRoot}~\cite{gu2024kpiroot} leverages Symbolic Aggregate Approximation (SAX) to create compact string representations of KPI time-series, allowing for fast similarity and causality computations via Jaccard similarity and Granger causality tests.
These symbolic approaches reduce computation time by 56.9\% compared to baselines while maintaining diagnostic accuracy.
\textit{Log3C}~\cite{he2018identifying} addresses the challenge of massive log volumes through a novel "Cascading Clustering" algorithm that avoids the quadratic complexity of traditional clustering by iteratively sampling, clustering, and matching log sequences, enabling the analysis of terabytes of daily logs in minutes.

\subsubsection{Low-Complexity Analytical Algorithms}

Several works introduce novel algorithms specifically designed for computational efficiency.
\textit{ShapleyIQ}~\cite{li2023shapleyiq} makes the typically exponential-time Shapley value computation feasible for real-time use by introducing a "splitting invariance" property that decomposes the problem, reducing complexity from exponential to $O(n log n)$.
This enables the system to analyze complex requests with hundreds of spans within milliseconds.
\textit{Minesweeper}~\cite{murali2021scalable} applies the efficient PrefixSpan algorithm for sequential pattern mining on app telemetry traces, completing analysis of tens of thousands of reports in under 3 minutes.

For root cause localization, several systems deliberately avoid computationally expensive operations.
\textit{PDiagnose}~\cite{hou2021Pdiagnose} eschews the costly construction of dependency graphs, instead employing lightweight unsupervised algorithms (KDE and WMA for anomaly detection) combined with a simple voting scheme, achieving polynomial time complexity suitable for real-time diagnosis.
\textit{$\epsilon$-Diagnosis}~\cite{shan2019diagnosis} tackles small-window long-tail latency by framing RCA as a two-sample hypothesis test using e-statistics based on energy distance correlation, enabling rapid analysis within seconds even for extremely short time windows.
\textit{FluxRank}~\cite{liu2019fluxrank} uses Kernel Density Estimation to quantify KPI changes and employs DBSCAN clustering with Pearson correlation for digest distillation, reducing localization time by over 80\% compared to manual approaches.

\subsubsection{Lightweight Model-Based Approaches}

Efficiency can also be achieved through careful model selection.
\textit{SLIM}~\cite{ren2024slim} generates interpretable rule sets to handle imbalanced fault data, using an efficient minorize-maximization approach for rule selection that incurs only about 15\% of the training overhead of state-of-the-art deep learning methods while maintaining superior accuracy.
\textit{MonitorRank}~\cite{kim2013root} splits computation into an intensive offline batch-mode engine for call graph generation and pseudo-anomaly clustering, and a lightweight real-time engine that performs a personalized PageRank-style random walk with time complexity O(Nc|V| + |E|), enabling rapid online diagnosis.

\subsection{Architectural Acceleration}
\label{subsec:architectural_acceleration}

The third strategy involves architectural patterns that accelerate computation through system design, particularly for large-scale data processing and dynamic environments.
These methods focus on how the system is organized to handle load, adapt to change, and leverage available computational resources.

\subsubsection{Parallel and Distributed Processing}

Parallel processing is essential for handling massive data volumes without sacrificing speed.
\textit{TraceContrast}~\cite{zhang2024trace} implements its core contrast sequential pattern mining algorithm on Apache Spark, distributing the computational load across a cluster to efficiently process large-scale trace data while maintaining real-time responsiveness.
\textit{FacGraph}~\cite{lin2018facgraph} similarly develops a distributed version of its frequent subgraph mining algorithm using MapReduce, significantly improving performance and scalability.
For causal analysis, \textit{Microscope}~\cite{lin2018microscope} parallelizes the computationally intensive PC algorithm for causality graph construction, while \textit{Sage}~\cite{gan2021sage} incorporates parallel training capabilities for its Graph Variational Autoencoder (GVAE) and Causal Bayesian Network (CBN) components, enabling the system to handle large-scale microservice architectures while maintaining model accuracy.
Both \textit{Murphy}~\cite{harsh2023murphy} and \textit{CIRCA}~\cite{li2022causal} incorporate architecture-aware parallelization strategies that adapt to available computational resources and workload characteristics.

\subsubsection{Incremental and Online Learning}

A critical architectural pattern for maintaining real-time performance in dynamic environments is incremental learning, which avoids costly full-model retraining.
\textit{CORAL}~\cite{wang2023incremental} is designed for online, near-real-time performance through two main features: an automatic trigger point detection module that initiates analysis early, and an incremental disentangled causal graph learning approach that efficiently updates the causal graph by decoupling state-invariant and state-dependent information.
This enables the system to adapt to new faults without starting from scratch.
Similarly, \textit{MRCA}~\cite{wang2024mrca} employs reinforcement learning (Q-learning) to dynamically terminate the causal graph construction process, learning an optimal policy to stop expansion when the graph is sufficient for accurate diagnosis, thereby significantly reducing end-to-end analysis time.

\subsubsection{Multi-Stage and Hierarchical Architectures}

Several systems achieve speed through carefully designed multi-stage pipelines that separate expensive offline computation from lightweight online analysis.
\textit{ChangeRCA}~\cite{yu2024changerca} is explicitly designed for speed to minimize MTTR, with a multi-stage design starting with a fast check for common canary-related issues.
Evaluation shows it can locate 90\% of defective changes in under 3 minutes, a 90\% reduction compared to baseline approaches.
\textit{CloudPD}~\cite{sharma2013cloudpd} achieves real-time performance through a layered, two-phase methodology where a computationally inexpensive Event Generation Engine first filters out normal intervals, and a more expensive correlation-based analysis is invoked only for suspicious intervals, enabling diagnosis within tens of seconds.
\textit{FChain}~\cite{nguyen2013fchain} similarly completes fault localization within a few seconds through lightweight monitoring and efficient change-point selection, while \textit{Roots}~\cite{jayathilaka2017performance} processes data asynchronously and periodically, demonstrating detection and diagnosis within minutes.

\subsubsection{Efficient Integration Frameworks}

Modern systems increasingly integrate multiple data sources and analytical components through efficient architectural patterns.
\textit{UniDiag}~\cite{zhang2024no} achieves an average online diagnosis time of under one second by separating computationally intensive offline training (graph construction, embedding, clustering) from a lightweight online diagnosis phase (embedding and distance comparison).
\textit{Groot}~\cite{wang2021groot} demonstrates end-to-end RCA completion in less than 5 seconds by constructing fine-grained event causality graphs with customizable rules and applying a customized PageRank algorithm.
\textit{MicroDig}~\cite{tao2024diagnosing} reduces diagnosis time from tens of minutes to under a minute by first identifying a small sub-graph of "association calls" to significantly prune the search space before performing more complex analysis.

\subsubsection{Specialized Approaches for Distinct Scenarios}

Several approaches target specific diagnostic scenarios with tailored performance optimizations.
For multi-dimensional analysis, \textit{ModelCoder}~\cite{cai2021modelcoder} analyzes high-level inter-service call data instead of detailed intra-service events, achieving localization within 80 seconds on average.
For trace-based analysis, \textit{TraceRank}~\cite{yu2023tracerank} combines lightweight spectrum analysis with PageRank-based random walk as a scalable alternative to deep learning models, while \textit{TraceStream}~\cite{zhou2023tracestream} employs lightweight trace embedding (TDTV) and non-iterative centrality-based localization, executing in milliseconds.
For change-related diagnosis, active RCA frameworks like those proposed in~\cite{ikeuchi2018root} use Greedy Entropy Minimization (GEM) and Reinforcement Learning to select and execute only the most informative actions in an optimal order, minimizing diagnostic time.

\subsection{LLM-Enhanced Efficiency Strategies}
\label{subsec:llm_efficiency}

The emergence of Large Language Models (LLMs) in RCA has introduced novel efficiency challenges and solutions.
These systems must address the unique computational constraints of LLM-based reasoning while maintaining real-time performance.

\subsubsection{Context Window and Token Optimization}

A primary challenge in LLM-based RCA is managing the massive volume of telemetry data that often exceeds LLM context windows.
\textit{KnowledgeMind}~\cite{ren2025multi} addresses token consumption through a service-by-service exploration strategy based on Monte Carlo Tree Search (MCTS), dramatically reducing the amount of information fed into the LLM in a single inference step and making the approach scalable to larger microservice systems.
\textit{OpenRCA}~\cite{xu2025openrca} introduces an RCA-agent architecture that uses code execution for data processing, avoiding costly token consumption by having the LLM generate and execute Python code to analyze telemetry data programmatically rather than processing raw data in the context window.
\textit{XPERT}~\cite{sun2025trioxpert} significantly reduces end-to-end latency by automating query authoring, generating domain-specific language queries in seconds through in-context learning, thereby eliminating the time-consuming manual query construction process.

\subsubsection{Model Optimization and Fine-Tuning}

Another line of work focuses on making LLM-based RCA cost-effective through model optimization.
\textit{eARCO}~\cite{goel2025earco} focuses on improving efficiency and cost-effectiveness by automatically optimizing prompts using PromptWizard and demonstrating that fine-tuned Small Language Models (SLMs), when paired with optimized prompts, can serve as a computationally efficient alternative to large expensive LLMs, reducing inference costs while maintaining performance.
\textit{ThinkFL}~\cite{zhang2025thinkfl} uses a lightweight LLM backbone (<10B parameters) with an efficient "Recursion-of-Thought" reasoning framework and progressive reinforcement fine-tuning, reducing end-to-end localization latency from minutes to seconds and making LLM-based RCA practical for production environments.

\section{GOAL 5: INTERPRETABILITY}\label{sec:goal5_interpretability}

As established in Section~\ref{sec:preliminaries}, a core objective of RCA is not only to identify the root cause (the "what") but also to explain the failure propagation path (the "how" and "why").
The goal of interpretability is to make these diagnostic results understandable, trustworthy, and verifiable for human operators.
This section organizes existing approaches by the \textit{strategy} they employ to achieve this goal.
We identify four primary strategies: 1) \textbf{Structural Interpretability}, which directly materializes the propagation graph; 2) \textbf{Semantic Interpretability}, which translates findings into human-readable narratives; 3) \textbf{Evidence-based and Rule-based Interpretability}, which exposes the underlying logic; and 4) \textbf{Interactive Interpretability}, which facilitates human-led exploration of the results.

\begin{table*}[t]
    \centering
    \caption{Overview of Interpretability Paradigms and Representative Works for RCA.}
    \label{tab:interpretability_summary}
    \footnotesize
    \setlength{\tabcolsep}{4pt}
    \renewcommand{\tabularxcolumn}[1]{m{#1}}
    \begin{tabularx}{\textwidth}{@{}>{\centering\arraybackslash}m{0.22\textwidth}>{\centering\arraybackslash}m{0.25\textwidth}X@{}}
        \toprule
        \textbf{Paradigm} & \textbf{Strategy} & \textbf{Papers} \\
        \midrule
        \multirow{3}{=}{\textbf{Structural Interpretability}} & Causal Graph Learning from Telemetry & CloudRanger~\cite{wang2018cloudranger}, LOUD~\cite{mariani2018localizing}, Sieve~\cite{thalheim2017sieve}, CauseRank~\cite{lu2022generic}, RUN~\cite{lin2024root}, AERCA~\cite{han2025root}, DyCause~\cite{pan2021faster}, CIRCA~\cite{li2022causal}, FlowRCA~\cite{wu2024flowrca}, RCSF~\cite{wang2015methodology} \\
        \cmidrule(lr){2-3}
        & Incorporating Domain Knowledge & GrayScope~\cite{zhang2024illuminating}, TS-InvarNet~\cite{hu2022ts}, HRLHF~\cite{wang2023root}, Atlas~\cite{xie2024cloud}, RealTCD~\cite{li2024realtcd} \\
        \cmidrule(lr){2-3}
        & Enriching Graph Semantics & MicroDig~\cite{tao2024diagnosing}, REASON~\cite{wang2023interdependent}, ICWS’17~\cite{jia2017approach}, Murphy~\cite{harsh2023murphy}, Chain-of-Event~\cite{yao2024chain} \\
        \midrule
        \multirow{3}{=}{\textbf{Semantic Interpretability}} & Natural Language Report Generation & RCACopilot~\cite{chen2024automatic}, SCELM~\cite{sun2025multimodal}, COCA~\cite{li2025coca}, SynergyRCA~\cite{xiang2025simplifying}, LM-PACE~\cite{zhang2024lm} \\
        \cmidrule(lr){2-3}
        & Interpretable Reasoning Process & Roy et al.~\cite{roy2024exploring}, OpenRCA~\cite{xu2025openrca}, Flow-of-Action~\cite{pei2025flow}, KnowledgeMind~\cite{ren2025multi}, TrioXpert~\cite{sun2025trioxpert}, mABC~\cite{zhang2024mabc} \\
        \cmidrule(lr){2-3}
        & High-Level Abstractions & Minesweeper~\cite{murali2021scalable}, COMET~\cite{wang2024large}, DéjàVu~\cite{li2022actionable} \\
        \midrule
        \multirow{3}{=}{\textbf{Evidence-based \& Rule-based Interpretability}} & Attribution to Evidence & Nezha~\cite{yu2023nezha}, LoFI~\cite{huang2024demystifying}, FluxRank~\cite{liu2019fluxrank}, ART~\cite{sun2024art}, TraceAnomaly~\cite{liu2020unsupervised}, PDiagnose~\cite{hou2021Pdiagnose} \\
        \cmidrule(lr){2-3}
        & Quantitative Attribution & ShapleyIQ~\cite{li2023shapleyiq}, CD-RCA~\cite{yokoyama2024causal}, LADRA~\cite{lu2017log}, DeepHunt~\cite{sun2025interpretable}, GAMMA~\cite{somashekar2024gamma} \\
        \cmidrule(lr){2-3}
        & Explicit Rule Generation & SLIM~\cite{ren2024slim}, CMDiagnostor~\cite{yu2023cmdiagnostor}, PatternMatcher~\cite{wu2021identifying}, KPIRoot~\cite{gu2024kpiroot}, Graphbasedrca~\cite{brandon2020graph}, Roots~\cite{jayathilaka2017performance}, DiagMLP~\cite{gao2025gnns} \\
        \midrule
        \multirow{2}{=}{\textbf{Interactive Interpretability}} & Visual Exploration & Zhou et al.~\cite{zhou2018fault}, Groot~\cite{wang2021groot} \\
        \cmidrule(lr){2-3}
        & Hypothesis-driven Investigation & EXPLAINIT!~\cite{jeyakumar2019explainit}, TraceDiag~\cite{ding2023tracediag}, ThinkFL~\cite{zhang2025thinkfl} \\
        \bottomrule
    \end{tabularx}
\end{table*}

\subsection{Structural Interpretability through Causal Graph Construction}
Structural interpretability aims to construct an explicit model of failure propagation, typically in the form of a graph.
This approach directly addresses the challenge of explaining the "how" and "why" of an incident by visualizing the causal chain from the root cause to the observed symptoms.
The graph itself becomes the explanation, providing a logical and verifiable narrative for operators.

A primary strategy in this area is to learn a causal graph from system telemetry.
Many methods employ statistical techniques, such as Granger causality, on performance metrics to infer a directed graph representing influence or dependency.
For instance, \textit{CloudRanger}~\cite{wang2018cloudranger}, \textit{LOUD}~\cite{mariani2018localizing}, \textit{Sieve}~\cite{thalheim2017sieve}, and \textit{CauseRank}~\cite{lu2022generic} all construct causal graphs from metrics and then apply ranking algorithms to pinpoint the most central nodes in the failure propagation.
\textit{RUN}~\cite{lin2024root} and \textit{AERCA}~\cite{han2025root} specifically adapt neural Granger causality for this task, capturing more complex temporal dependencies.
\textit{DyCause}~\cite{pan2021faster} further extends this to discover time-varying causalities, showing how relationships evolve during an incident.
Other methods like \textit{CIRCA}~\cite{li2022causal} and \textit{FlowRCA}~\cite{wu2024flowrca} ground their graph construction in formal causal inference theory, identifying root causes as "interventions" that break the learned normal causal relationships.

To improve the accuracy and plausibility of these graphs, some approaches incorporate domain knowledge.
\textit{GrayScope}~\cite{zhang2024illuminating} and \textit{TS-InvarNet}~\cite{hu2022ts} refine data-driven graphs by starting with an expert-defined "causality skeleton."
\textit{HRLHF}~\cite{wang2023root} uses reinforcement learning to actively query human experts, efficiently integrating their knowledge into the graph discovery process.
\textit{Atlas}~\cite{xie2024cloud} and \textit{RealTCD}~\cite{li2024realtcd} leverage Large Language Models (LLMs) to parse documentation and generate a high-quality prior causal structure.

Other approaches focus on enriching the graph's semantics.
\textit{MicroDig}~\cite{tao2024diagnosing} constructs a heterogeneous graph distinguishing between services and calls, while \textit{REASON}~\cite{wang2023interdependent} models interdependent networks across system layers (\eg pods and servers).
\textit{ICWS’17}~\cite{jia2017approach} builds a two-layer graph modeling both inter-service topology and intra-service control flow.
\textit{Murphy}~\cite{harsh2023murphy} uses a Markov Random Field to handle cyclic dependencies, which are common in real systems but problematic for many causal discovery algorithms.
\textit{Chain-of-Event}~\cite{yao2024chain} automatically learns a weighted event-causal graph where parameters have intuitive physical meanings, allowing engineers to inspect the model's reasoning.
Finally, \textit{RCSF}~\cite{wang2015methodology} focuses on generating fault propagation sequences even with incomplete monitoring coverage.

\subsection{Semantic Interpretability through Natural Language and High-Level Concepts}
Semantic interpretability focuses on translating technical findings into human-readable narratives or high-level concepts.
This strategy has gained significant traction with the advent of LLMs, which excel at synthesizing complex information into coherent text and classifying issues into understandable categories.

A prominent application of LLMs is generating natural language reports.
\textit{RCACopilot}~\cite{chen2024automatic} and \textit{SCELM}~\cite{sun2025multimodal} analyze diagnostic data to produce structured reports that summarize the incident, identify the root cause, and suggest solutions.
To ground the LLM's generation in factual data, many approaches adopt Retrieval-Augmented Generation (RAG).
\textit{COCA}~\cite{li2025coca} retrieves relevant source code, while \textit{SynergyRCA}~\cite{xiang2025simplifying} queries a real-time "StateGraph" of the system.
To enhance trust, \textit{LM-PACE}~\cite{zhang2024lm} provides a calibrated confidence score for the LLM's output.

Another line of work focuses on making the LLM's reasoning process itself interpretable.
Frameworks like \textit{Roy et al.}~\cite{roy2024exploring} and \textit{OpenRCA}~\cite{xu2025openrca} use a "Chain-of-Thought" or "ReAct" paradigm, where the LLM externalizes its diagnostic steps.
Multi-agent systems like \textit{Flow-of-Action}~\cite{pei2025flow}, \textit{KnowledgeMind}~\cite{ren2025multi}, \textit{TrioXpert}~\cite{sun2025trioxpert}, and \textit{mABC}~\cite{zhang2024mabc} structure this reasoning process further, assigning specific sub-tasks to specialized agents to make the overall analysis more robust and transparent.

Beyond LLMs, other methods provide semantic meaning through high-level abstractions.
\textit{Minesweeper}~\cite{murali2021scalable} discovers and ranks sequential event patterns (\eg "PlayVideo -> DeleteStory") that are distinctive to buggy sessions, providing a clear narrative of user actions leading to a failure.
\textit{COMET}~\cite{wang2024large} uses an LLM to extract keywords from logs, which helps engineers quickly understand the nature of an incident.
\textit{DéjàVu}~\cite{li2022actionable} provides interpretability by classifying a failure into a known category and retrieving a similar historical incident, explaining the current problem by analogy.

\subsection{Evidence-based and Rule-based Interpretability}
This category of methods achieves interpretability by making the logic of the diagnosis explicit, either by exposing the underlying evidence or by presenting the conclusion as a set of human-readable rules.
This approach builds trust by allowing operators to understand and verify the "why" behind a conclusion.

Several methods provide interpretability by attributing a finding to specific, understandable evidence.
\textit{Nezha}~\cite{yu2023nezha} explains a failure by showing the deviation between "expected" and "actual" execution patterns.
\textit{LoFI}~\cite{huang2024demystifying} extracts specific fault-indicating phrases from logs, directly answering "what went wrong." 
\textit{FluxRank}~\cite{liu2019fluxrank} distills thousands of alerts into a few "digests" (\eg "27 machines in module M1 experienced CPU overload").
\textit{ART}~\cite{sun2024art} uses a "unified failure representation" where each feature dimension directly corresponds to an original data channel, showing which signals are deviating.
\textit{TraceAnomaly}~\cite{liu2020unsupervised} uses a handcrafted vector where each dimension represents a specific (service, callpath) tuple, making the source of deviation clear.
\textit{PDiagnose}~\cite{hou2021Pdiagnose} provides concrete evidence by outputting the specific KPI names and raw log entries that are anomalous.

Quantitative attribution is another powerful technique.
\textit{ShapleyIQ}~\cite{li2023shapleyiq} and \textit{CD-RCA}~\cite{yokoyama2024causal} use Shapley values to quantify precisely how much each component contributed to the failure.
\textit{LADRA}~\cite{lu2017log} provides a probabilistic diagnosis of resource contention (CPU, memory, etc.) to explain why a task is slow.
\textit{DeepHunt}~\cite{sun2025interpretable} calculates an interpretable score based on an instance's own anomaly and the anomalies of its neighbors in the propagation path.
\textit{GAMMA}~\cite{somashekar2024gamma} uses feature-omission studies to explain the type of bottleneck (\eg CPU-bound) by observing performance drops when certain metrics are excluded.

Finally, some methods generate explicit, human-readable rules.
\textit{SLIM}~\cite{ren2024slim} produces a set of decision rules in Disjunctive Normal Form (\eg \texttt{IF cpu\_usage > 80 THEN fault}).
\textit{CMDiagnostor}~\cite{yu2023cmdiagnostor} uses rule-based pruning and ranking keys, providing a clear rationale for its choices.
\textit{PatternMatcher}~\cite{wu2021identifying} classifies anomalies into physically meaningful patterns before ranking, making the reasoning transparent.
\textit{KPIRoot}~\cite{gu2024kpiroot} uses a two-factor logic (''it looks similar and it happened first'') that is intuitive for operators.
\textit{Graphbasedrca}~\cite{brandon2020graph} matches an anomaly to a library of pre-labeled, human-understandable patterns.
\textit{Roots}~\cite{jayathilaka2017performance} employs a combination of statistical methods to identify the bottleneck, justifying its approach through a majority vote.
The study \textit{DiagMLP}~\cite{gao2025gnns} provides interpretability for an entire class of models by showing that their performance stems from data fusion rather than complex GNNs, clarifying the true drivers of success.

\subsection{Interactive Interpretability through Visual and Exploratory Interfaces}
Interactive interpretability empowers operators by providing tools to visually explore data, test hypotheses, and engage directly with the findings.
This approach facilitates a dialogue between the operator and the system, where the operator's domain knowledge can guide the investigation.

Visual exploration of system behavior is a cornerstone of this approach.
The empirical study in \textit{Zhou et al.}~\cite{zhou2018fault} validates this, showing that visual trace analysis significantly helps developers understand fault propagation.
Systems like \textit{Groot}~\cite{wang2021groot} provide interactive interfaces that allow operators to click on graph nodes for details, filter the view, and trace failure paths, transforming an abstract graph into a concrete investigative tool.

Other systems facilitate hypothesis-driven investigation.
\textit{EXPLAINIT!}~\cite{jeyakumar2019explainit} offers a declarative, SQL-like interface for operators to formulate and test causal hypotheses against time-series data.
This allows operators to leverage their domain expertise to guide the analysis.
\textit{TraceDiag}~\cite{ding2023tracediag} learns an interpretable "filtering tree" (a form of decision tree) that explains its reasoning for pruning the search space, allowing engineers to understand the system's focus.
\textit{ThinkFL}~\cite{zhang2025thinkfl} uses a "Recursion-of-Thought" mechanism that allows an LLM to dynamically query data tools, with the entire reasoning path being transparent and verifiable by SREs.
By turning RCA into an interactive and exploratory process, these methods bridge the gap between automated analysis and human-led problem-solving.
\section{GOAL 6: MULTI-GRANULARITY}\label{sec:goal6_multigranularity}

As established in our formalization (Section~\ref{sec:preliminaries}), the primary objective of most RCA systems is to identify the root cause event node(s) $r$ within the incident propagation graph $\mathcal{G}$.
The goal of multi-granularity directly addresses the challenge of localizing this root cause at varying levels of abstraction.
This capability is essential because different roles in the incident management lifecycle require diagnoses at different depths.
Site Reliability Engineers (SREs) need coarse-grained localization for rapid mitigation (reducing MTTR), while developers require fine-grained analysis to implement permanent fixes (improving MTBF).
The achievable precision of the output is fundamentally constrained by the granularity of the input observation space $\mathcal{O}$.
This section, therefore, reviews approaches based on their ability to provide outputs at hierarchical levels of abstraction, from the service level down to the code.

\begin{table*}[t]
    \centering
    \caption{Overview of Multi-Granularity Paradigms and Representative Works for RCA.}
    \label{tab:multigranularity_summary}
    \footnotesize
    \setlength{\tabcolsep}{4pt}
    \renewcommand{\tabularxcolumn}[1]{m{#1}}
    \begin{tabularx}{\textwidth}{@{}>{\centering\arraybackslash}m{0.25\textwidth}>{\centering\arraybackslash}m{0.25\textwidth}X@{}}
        \toprule
        \textbf{Paradigm} & \textbf{Strategy} & \textbf{Papers} \\
        \midrule
        \multirow{5}{=}{\textbf{Hierarchical Drill-Down}} & From Service to Infrastructure/Instance & FAMOS~\cite{duan2025famos}, FaaSRCA~\cite{huang2024faasrca}, HALO~\cite{zhang2021halo}, KnowledgeMind~\cite{ren2025multi}, MicroIRC~\cite{zhu2024microirc}, REASON~\cite{wang2023interdependent}, SwissLog~\cite{li2022swisslog} \\
        \cmidrule(lr){2-3}
        & From Service to Component/Metric & AERCA~\cite{han2025root}, CausalRCA~\cite{xin2023causalrca}, CauseInfer~\cite{chen2014causeinfer}, CloudRCA~\cite{zhang2021cloudrca}, CMDiagnostor~\cite{yu2023cmdiagnostor}, FlowRCA~\cite{wu2024flowrca}, HeMiRCA~\cite{zhu2024hemirca}, ICSOC'20~\cite{wu2020performance}, LatentScope~\cite{xie2024microservice}, MRCA~\cite{wang2024mrca} \\
        \cmidrule(lr){2-3}
        & From Service to Code/Change & ChangeRCA~\cite{yu2024changerca}, COCA~\cite{li2025coca}, LogFaultFlagger~\cite{amar2019mining}, MEPFL~\cite{zhou2019latent}, Nezha~\cite{yu2023nezha}, Raccoon~\cite{zhao2023identifying}, ServerRCA~\cite{shi2023serverrca}, TrinityRCL~\cite{gu2023trinityrcl} \\
        \midrule
        \multirow{3}{=}{\textbf{Multi-Dimensional \& Fine-Grained Localization}} & Multi-Attribute Pattern Mining & TraceContrast~\cite{zhang2024trace}, TVDiag~\cite{xie2025tvdiag} \\
        \cmidrule(lr){2-3}
        & Operation \& Span-Level Analysis & faultstudy~\cite{zhou2018fault}, SpanGraph~\cite{kong2024enhancing}, TraceNet~\cite{yang2023tracenet} \\
        \bottomrule
    \end{tabularx}
\end{table*}

\subsection{Hierarchical Granularity Levels in RCA}

The granularity of root cause identification is fundamentally constrained by two factors: the granularity of the input observational data and the effectiveness of the inference method in utilizing that data.
As illustrated in our formalization (Section\ref{sec:preliminaries}), the observation space $\mathcal{O}$ contains telemetry data at various abstraction levels, and the analysis can only be as fine-grained as the most detailed available data permits.


The pursuit of multi-granularity in RCA is driven by the diverse needs of different roles within the incident management lifecycle.
While SREs often require rapid, coarse-grained localization (\eg identifying a faulty service or instance) to facilitate immediate mitigation, developers need fine-grained, deep localization (\eg pinpointing a specific metric, code change, or function) to implement permanent fixes.
An effective RCA system must therefore provide outputs at multiple, hierarchical levels of abstraction.
We review approaches based on the depth and precision of their localization capabilities, categorizing them by their ability to drill down from the service level to the infrastructure, component, and code levels.

\subsubsection{From Service to Infrastructure and Component Granularity}

A primary objective in multi-granularity RCA is to bridge the gap between service-level symptoms and their underlying causes within the infrastructure or application components.
Several approaches achieve this by constructing hierarchical models that explicitly connect different system layers.

\textbf{Two-Stage Top-Down Localization.}
A common strategy involves a two-stage, top-down localization process that first identifies a faulty service and then drills down to pinpoint a more specific root cause.
\textit{CauseInfer}~\cite{chen2014causeinfer} employs a two-layered hierarchical causality graph: a coarse-grained service dependency graph constructed by analyzing network traffic delays, and for each service, a fine-grained metric causality graph built using the PC-algorithm.
When an SLO violation occurs, the system first traverses the service graph to localize the faulty service, then traverses the metric graph using depth-first search with CUSUM-based change detection to identify anomalous root cause metrics.
Similarly, the method proposed in ICSOC'20~\cite{wu2020performance} constructs a service dependency graph and uses Personalized PageRank to identify potential culprit services, then applies an autoencoder-based model trained on normal data to analyze reconstruction errors of live metrics, pinpointing the root cause at the metric-level within each candidate service.
\textit{MRCA}~\cite{wang2024mrca} extends this paradigm by fusing features from logs and traces for more accurate anomaly detection and initial ranking of abnormal services, then performing fine-grained causal analysis using Granger causality on the metrics of top-ranked services, with a Q-learning agent dynamically terminating graph expansion to balance accuracy and speed.
\textit{KnowledgeMind}~\cite{ren2025multi} employs a Monte Carlo Tree Search (MCTS) process guided by LLMs to identify the faulty service through service-by-service reasoning, followed by a dedicated Service-Pod Agent that drills down to the specific faulty pod.

Several methods refine the metric-level localization through sophisticated anomaly correlation techniques.
\textit{HeMiRCA}~\cite{zhu2024hemirca} leverages the monotonic correlation between heterogeneous monitoring data by constructing span vectors from traces to represent invocation latency, using a Variational Autoencoder (VAE) to compute anomaly scores, and calculating Spearman rank correlation with individual metric time series to rank suspicious metrics and their corresponding services.
\textit{CausalRCA}~\cite{xin2023causalrca} applies gradient-based causal structure learning (DAG-GNN) to build a weighted directed acyclic graph representing causal dependencies between metrics, then applies PageRank to identify root causes at both service and metric levels without strict distributional assumptions.
\textit{FlowRCA}~\cite{wu2024flowrca} constructs a metric-level causality graph using normalizing flows to infer causal direction and quantifies causal impacts using Conditional Average Treatment Effect (CATE), applying Personalized PageRank on the inverted anomalous subgraph to rank metrics.
\textit{AERCA}~\cite{han2025root} uses an autoencoder-based framework integrating Granger causal discovery, defining anomalies as interventions on exogenous variables to identify not only the root-cause time series but also the specific time steps of the intervention.

\textbf{Unified Multi-Layer Graph Models.}
Other approaches build unified, multi-layered graphs that inherently represent the system's hierarchical structure.
\textit{HALO}~\cite{zhang2021halo} automatically learns an Attribute Hierarchy Graph (such as Node $\rightarrow$ Cluster $\rightarrow$ Datacenter) by analyzing pairwise conditional entropy between attributes, then uses a failure-aware random walk to generate promising search paths and performs a self-adaptive top-down search with OTSU-based pruning to find the optimal fault-indicating attribute-value combination at the appropriate granularity.
\textit{REASON}~\cite{wang2023interdependent} models the system as an interdependent network of high-level servers and low-level pods, employing a hierarchical Graph Neural Network (GNN) to learn intra-level and inter-level causal relationships, combining topological causal discovery with individual causal discovery based on Extreme Value Theory.
\textit{FaaSRCA}~\cite{huang2024faasrca} constructs a "Global Call Graph" that integrates multi-modal observability data from both application functions and platform components (such as Kubernetes pods), using an unsupervised Graph Attention Network (GAT) based autoencoder trained on normal operations to compare reconstruction errors and identify root causes across the full serverless lifecycle.
\textit{MicroIRC}~\cite{zhu2024microirc} builds a Heterogeneous Weighted Topology (HWT) graph of services, instances, and hosts, running a personalized random walk to generate root cause candidates and feeding them with real-time metrics into a pre-trained GNN model (MetricSage) to produce a final ranked list at the instance level.
\textit{SwissLog}~\cite{li2022swisslog} provides multi-granularity localization by first detecting anomalies at the execution instance level from interleaved logs. 
It then uses a pre-constructed ID relation graph, which maps dependencies between different entity types (such as application, container, and block), to pinpoint the specific anomalous instance, allowing operators to drill down from a system-wide issue to a fine-grained root cause.

\textbf{Specialized Domain Integration.}
Some methods achieve multi-granularity through domain-specific integration strategies.
\textit{CMDiagnostor}~\cite{yu2023cmdiagnostor} operates across multiple granularities by ingesting fine-grained, method-level Call Metric Data and constructing an ambiguity-free call graph using a novel regression-based method (AmSitor) to resolve upstream-downstream call correspondences, ultimately outputting a ranked list of coarse-grained services.
\textit{CloudRCA}~\cite{zhang2021cloudrca} integrates heterogeneous data sources (including KPIs, logs, and topology) into a Knowledge-informed Hierarchical Bayesian Network (KHBN) with a hierarchical root cause layer, enabling the model to pinpoint both the high-level faulty module and the specific, low-level fault type.
\textit{FAMOS}~\cite{duan2025famos} collects metrics from both host and container levels and correlates them with service-level traces using a late-fusion paradigm with Gaussian-attention and cross-attention mechanisms, enabling it to identify fine-grained root cause types such as "Host CPU overload" or "Container stopped".
\textit{LatentScope}~\cite{xie2024microservice} models heterogeneous root cause candidates (including services, pods, hosts, databases, and software changes) as latent variables in a dual-space graph, using a Regression-based Latent-space Intervention Recognition (RLIR) algorithm to infer anomalous latent variables even with limited observability.

\subsubsection{From Service to Code-Level Granularity}

The ultimate goal for many RCA systems is to provide code-level localization, directly guiding developers to the source of a fault.
This requires sophisticated techniques that can connect high-level system behavior to specific code artifacts.

\textbf{Direct Code Integration.}
Several methods achieve code-level localization by integrating code-related information directly into their analysis.
\textit{TrinityRCL}~\cite{gu2023trinityrcl} constructs a heterogeneous causal graph containing services, hosts, metrics, and faults (which represent code exceptions extracted from logs), assigning anomaly scores to edges using correlation algorithms (such as DTW and CORT) and using Random Walk with Restart (RWR) to simulate anomaly propagation and identify root causes across application, service, host, metric, and code levels.
\textit{Nezha}~\cite{yu2023nezha} transforms multi-modal data (including metrics, logs, and traces) into a unified stream of events structured into event graphs, comparing the frequency of event patterns (which are represented as subgraphs) between normal and faulty periods to identify deviating patterns that correspond to code regions or resource types, providing interpretable fine-grained root cause candidates.
\textit{COCA}~\cite{li2025coca} leverages LLMs to analyze issue reports, using static analysis-based backtracking to link log messages to source code locations, building a call graph patched with a novel RPC bridging method to reconstruct execution paths, and combining issue reports with retrieved code snippets for root cause inference at the class and method level.

\textbf{Software Change Identification.}
Another category focuses on identifying specific software changes as the root cause.
\textit{ChangeRCA}~\cite{yu2024changerca} refines service-level RCA output through a three-stage framework: using cascaded Difference-in-Differences (DiD) to detect faulty canary releases, filtering non-change-related faults, and scoring recent changes by integrating KPIs, service dependency graphs, and change timing to pinpoint the specific defective software change.
\textit{Raccoon}~\cite{zhao2023identifying} bridges the semantic gap between user-reported incidents and software changes by representing incidents at a user-perceived functional level using Fault Trees and Software Product Lines, mining causal knowledge with a Tree GNN to build a knowledge base linking incidents to changes, and recommending root-cause changes at multiple granularities (ranging from product line to specific change).

\textbf{Test Log and Trace Analysis.}
Methods focusing on test environments and operational traces also achieve fine-grained localization.
\textit{LogFaultFlagger}~\cite{amar2019mining} localizes faults from entire test log files down to specific log lines by calculating a score combining line-level Inverse Document Frequency (line-IDF) with historical fault association, using an Exclusive K-Nearest Neighbors (EKNN) algorithm to predict product faults and flag the most probable cause lines.
\textit{MEPFL}~\cite{zhou2019latent} learns from system trace logs to predict latent errors and locate faulty microservices, extracting comprehensive features at both trace-level and microservice-level to train machine learning models that identify the fault type and responsible service.
\textit{ServerRCA}~\cite{shi2023serverrca} employs a hierarchical matching framework using contrastive learning on operating system logs, analyzing at three levels (namely Fault, Module, and Event) with BERT-based encoders to drill down from general symptoms to specific, actionable fault events, constructing a fault propagation chain via a knowledge graph.

\subsubsection{Multi-Dimensional and Fine-Grained Localization}

Beyond simple hierarchical localization, some advanced methods provide multi-dimensional analysis, identifying root causes as a combination of factors across different system dimensions.

\textbf{Multi-Attribute Pattern Mining.}
\textit{TraceContrast}~\cite{zhang2024trace} frames RCA as a contrast sequential pattern mining problem, representing traces as sequences of attribute sets (including service version, API route, and OS version) and applying a parallel contrast sequential pattern mining algorithm to find patterns frequent in anomalous paths but rare in normal ones, ranked using spectrum analysis (Ochiai) to provide precise, hierarchical diagnosis.
\textit{TVDiag}~\cite{xie2025tvdiag} builds an instance correlation graph and employs a GNN-based multimodal co-learning module with task-oriented supervised contrastive learning and cross-modal contrastive learning to simultaneously perform root cause localization (identifying the faulty instance) and failure type identification, serving both localization and classification needs.

\textbf{Operation and Span-Level Analysis.}
Methods that refine the definition of system entities achieve even finer granularity.
\textit{TraceNet}~\cite{yang2023tracenet} constructs a Service Dependency Graph at the operation level (which represents specific API endpoints) rather than the service level, quantifying microservice abnormality by distinguishing between inner-abnormality and outer-abnormality to handle propagation effects, enabling it to differentiate between business functions within a single microservice.
\textit{SpanGraph}~\cite{kong2024enhancing} operates at the span level by constructing a directed graph where nodes represent unique microservice requests (characterized by NodeId, InstanceId, ServiceName, and ApiName) and edges represent invocations, using a Graph Convolutional Network (GCN) to classify edges as normal or anomalous and localizing faults to the starting node of anomalous edges.
The empirical study presented in the fault analysis work~\cite{zhou2018fault} proposes two distinct trace visualization strategies: "Microservice as Node" (which provides service-level view) and "Microservice State as Node" (which provides state-level view), enabling developers to analyze failures at different abstraction levels.

\section{GOAL 7: ACTIONABILITY}\label{sec:goal7_actionability}

As established in Section~\ref{sec:preliminaries}, the core output of the RCA function is the incident propagation graph $\mathcal{G}$, which provides the diagnostic explanation of a failure.
However, a diagnosis alone is insufficient for effective incident management; it must be translated into concrete remedial actions.
Actionability addresses this need by focusing on translating diagnostic outputs into concrete operational directives to guide swift remediation.
This section examines how research transforms RCA into a prescriptive tool, organizing the discussion around three dimensions: direct remediation generation (Section~\ref{sec:remediation_guidance}), automated responsibility assignment (Section~\ref{sec:responsibility_assignment}), and actionable knowledge provision (Section~\ref{sec:actionable_knowledge}).

\begin{table*}[t]
    \centering
    \caption{Overview of Actionability Paradigms and Representative Works for RCA.}
    \label{tab:actionability_summary}
    \footnotesize
    \setlength{\tabcolsep}{4pt}
    \renewcommand{\tabularxcolumn}[1]{m{#1}}
    \begin{tabularx}{\textwidth}{@{}>{\centering\arraybackslash}m{0.22\textwidth}>{\centering\arraybackslash}m{0.25\textwidth}X@{}}
        \toprule
        \textbf{Paradigm} & \textbf{Strategy} & \textbf{Papers} \\
        \midrule
        \multirow{3}{=}{\textbf{Direct Remediation Generation}} & Automated Actuation \& Resource Management & CloudPD~\cite{sharma2013cloudpd}, Sage~\cite{gan2021sage} \\
        \cmidrule(lr){2-3}
        & LLM-Driven Remediation Planning & SynergyRCA~\cite{xiang2025simplifying}, SCELM~\cite{sun2025multimodal}, mABC~\cite{zhang2024mabc}, Ahmed et al.~\cite{ahmed2023recommending} \\
        \cmidrule(lr){2-3}
        & Failure Classification \& Knowledge-Based Remediation & Déjàvu~\cite{li2022actionable}, MEPFL~\cite{zhou2019latent}, AutoMAP~\cite{ma2020automap}, LogKG~\cite{sui2023logkg} \\
        \midrule
        \textbf{Automated Responsibility Assignment} & Automated Triage \& Escalation & RCACopilot~\cite{chen2024automatic}, COMET~\cite{wang2024large}, RCAgent~\cite{wang2024rcagent} \\
        \midrule
        \multirow{3}{=}{\textbf{Actionable Knowledge Provision}} & Precision Diagnostic Artifacts & DéjàVu~\cite{li2022actionable}, Xpert~\cite{sun2025trioxpert}, Ikeuchi et al.~\cite{ikeuchi2018root} \\
        \cmidrule(lr){2-3}
        & Contextual Knowledge Retrieval & RCAgent~\cite{wang2024rcagent}, Roy et al.~\cite{roy2024exploring}, RCACopilot~\cite{chen2024automatic}, ICLRCA~\cite{zhang2024automated}, SynergyRCA~\cite{xiang2025simplifying} \\
        \cmidrule(lr){2-3}
        & Human-in-the-Loop Validation \& Trust Calibration & Groot~\cite{wang2021groot}, HRLHF~\cite{wang2023root}, Chain-of-Event~\cite{yao2024chain}, TraceDiag~\cite{ding2023tracediag}, Nezha~\cite{yu2023nezha}, LM-PACE~\cite{zhang2024lm}, GMTA~\cite{guo2020graph}, Eadro~\cite{lee2023eadro} \\
        \bottomrule
    \end{tabularx}
\end{table*}

\subsection{Direct Remediation Generation}\label{sec:remediation_guidance}

The most direct form of actionability is achieved by systems that automatically generate or trigger specific repair actions, thereby closing the loop between detection and resolution with minimal human intervention.

\subsubsection{Automated Actuation and Resource Management}

Early work in this area integrated diagnostic frameworks with actuation controllers to enable automated corrective actions.
\textit{CloudPD}~\cite{sharma2013cloudpd} and \textit{Sage}~\cite{gan2021sage} exemplify this approach by implementing closed-loop systems that, upon identifying faults such as resource contention or performance bottlenecks, automatically trigger remedial operations.
These operations include VM reconfiguration, CPU frequency scaling, resource re-partitioning, or microservice scaling to restore quality of service.
By directly coupling diagnosis with actuation, these systems significantly reduce Mean Time to Recovery (MTTR) and minimize the need for manual operator intervention during critical incidents.

\subsubsection{LLM-Driven Remediation Planning}

The advent of Large Language Models has enabled a new paradigm of context-aware, nuanced remediation generation.
Rather than relying on predefined action templates, LLM-based systems can generate tailored repair plans that account for incident-specific context and organizational practices.

\textit{SynergyRCA}~\cite{xiang2025simplifying} demonstrates this capability in Kubernetes environments by producing not only diagnostic reports but also precise, executable commands (\eg \texttt{kubectl patch}) tailored to the specific incident.
This provides operators with validated, one-step solutions.
Similarly, \textit{SCELM}~\cite{sun2025multimodal} and \textit{mABC}~\cite{zhang2024mabc} incorporate dedicated solution generation modules or specialized agents (\eg ''Solution Engineer'' agents) that formulate explicit resolution steps as integral components of their analytical output.
The foundational work by Ahmed et al.~\cite{ahmed2023recommending} provided large-scale empirical evidence for this approach, demonstrating through evaluation on over 40,000 Microsoft incidents that fine-tuned LLMs can effectively recommend concrete mitigation steps alongside root causes, with their utility validated by human incident owners.

\subsubsection{Failure Classification and Knowledge-Based Remediation}

An alternative approach enhances actionability through failure type classification coupled with knowledge-based remediation databases.
Systems such as \textit{Déjàvu}~\cite{li2022actionable} and \textit{MEPFL}~\cite{zhou2019latent} maintain taxonomies of failure categories (\eg configuration-related, resource-related, bad requests, slow queries), with each category mapped to established remediation workflows.
This categorical approach enables immediate guidance on appropriate response measures and is particularly valuable in large organizations where different failure types require expertise from distinct teams or follow different escalation procedures.
However, the effectiveness of this approach depends critically on the comprehensiveness and adaptability of the underlying taxonomy to organization-specific contexts and evolving architectures.

A complementary technique leverages historical incident databases through case-based reasoning.
Methods such as \textit{Déjàvu}~\cite{li2022actionable}, \textit{AutoMAP}~\cite{ma2020automap}, and \textit{LogKG}~\cite{sui2023logkg} retrieve similar historical incidents based on symptom patterns and failure characteristics, surfacing not only technical details but also the remediation actions that were successfully applied, their effectiveness, and lessons learned.
This approach provides battle-tested solutions validated in production environments and captures organizational knowledge that may not be formally documented.
However, systems must account for architectural changes and software version updates that may invalidate past solutions.

\subsection{Automated Responsibility Assignment}\label{sec:responsibility_assignment}

In large, complex organizations, knowing \textit{what} to do is insufficient without determining \textit{who} should do it.
Misdirected incidents lead to significant delays as tickets are manually rerouted between teams.
Automated responsibility assignment addresses this challenge by routing incidents to appropriate personnel based on diagnosed root causes and organizational context.

\textit{RCACopilot}~\cite{chen2024automatic} and \textit{COMET}~\cite{wang2024large} directly tackle this problem through automated triage systems.
These frameworks analyze incident data to identify the appropriate on-call engineers or teams responsible for affected components.
\textit{RCACopilot} maintains comprehensive mappings between system components, failure types, and responsible teams, enabling nuanced routing where, for instance, a memory leak is directed to the development team while resource exhaustion in the same service escalates to the infrastructure team.
\textit{COMET} employs LLMs to extract keywords from incident data and matches them to historical incidents to determine team ownership, demonstrating a 30\% improvement in triage accuracy and 35\% reduction in Time to Mitigation at Microsoft.

Advanced systems implement dynamic escalation paths that adapt based on incident severity, resolution progress, and team availability.
If no progress occurs within defined time windows, incidents automatically escalate to senior engineers or involve additional teams.
\textit{RCAgent}~\cite{wang2024rcagent} integrates responsibility assignment as a core feature of its autonomous agent framework, making it practical for DevOps workflows.
The primary challenge lies in maintaining accurate mappings between technical components and organizational responsibilities in rapidly evolving environments, requiring either continuous manual maintenance or automated inference from operational data and organizational patterns.

\subsection{Actionable Knowledge Provision}\label{sec:actionable_knowledge}

Beyond fully automated repairs or assignments, another line of research focuses on empowering human operators by making the diagnostic process itself more action-oriented.
This is achieved by providing actionable knowledge or intermediate artifacts that accelerate subsequent manual investigation and resolution steps.

\subsubsection{Precision Diagnostic Artifacts}

One approach refines the diagnostic output to have clear operational implications.
Rather than identifying only a faulty component (too coarse) or individual metrics (too fine-grained), systems can provide actionable diagnostic units that directly suggest remediation classes.
\textit{DéjàVu}~\cite{li2022actionable} exemplifies this by identifying a ''failure unit''---a combination of a faulty component and an indicative metric group (\eg `high memory usage' on a specific Docker container).
This output immediately suggests categories of remedial actions (\eg investigate memory leaks, increase resource limits), bridging the gap between diagnosis and mitigation without requiring extensive interpretation.

Another powerful technique equips operators with precise investigation tools.
\textit{Xpert}~\cite{sun2025trioxpert} operationalizes this by using an LLM to translate natural language incident descriptions into precise, executable Kusto Query Language (KQL) queries.
This transforms a passive ticket into an active investigation, allowing engineers to immediately retrieve the necessary telemetry data without the cognitive load and potential errors of manual query construction.
Similarly, the active probing mechanism proposed by Ikeuchi et al.~\cite{ikeuchi2018root} generates more discriminative diagnostic data through targeted user action execution, enabling operators to perform precise, low-impact fixes (\eg restarting a single process) instead of resorting to broad, disruptive actions (\eg rebooting an entire device).

\subsubsection{Contextual Knowledge Retrieval}

Agent-based systems leverage tool-use and knowledge retrieval to ground their analysis in validated operational wisdom.
Frameworks such as \textit{RCAgent}~\cite{wang2024rcagent} and the LLM-based agent evaluated by Roy et al.~\cite{roy2024exploring} interact with live diagnostic services, knowledge base articles (KBAs), and historical incident databases.
By retrieving proven solutions from past incidents or internal documentation, these systems ground their recommendations in battle-tested practices rather than generic heuristics.
\textit{RCACopilot}~\cite{chen2024automatic} extends this by generating detailed explanatory narratives that bridge knowledge gaps for on-call engineers unfamiliar with specific components, providing not just technical findings but also business context and relationships.

Several recent approaches employ retrieval-augmented generation (RAG) to combine LLM capabilities with historical incident knowledge.
\textit{ICLRCA}~\cite{zhang2024automated} implements a standard RAG pipeline that retrieves similar historical cases and uses them to prompt LLMs for contextualized root cause analysis.
\textit{SynergyRCA}~\cite{xiang2025simplifying} constructs a graph-based RAG system using StateGraphs and MetaGraphs to capture runtime spatial-temporal relationships in Kubernetes, enabling more precise context retrieval.
These approaches ensure that generated recommendations are grounded in actual operational experience rather than generic knowledge, though they face challenges related to data quality, privacy, and the representativeness of historical cases in rapidly evolving environments.

\subsubsection{Human-in-the-Loop Validation and Trust Calibration}

While automation accelerates incident response, human expertise remains essential for handling novel failures and validating high-stakes decisions.
Several systems incorporate mechanisms for expert feedback integration and confidence assessment to enhance the trustworthiness of actionable outputs.

Systems like \textit{Groot}~\cite{wang2021groot}, \textit{HRLHF}~\cite{wang2023root}, and \textit{Chain-of-Event}~\cite{yao2024chain} enable human experts to inject domain knowledge by refining event relationships, validating causal dependencies, or adjusting analysis parameters.
\textit{HRLHF}~\cite{wang2023root} and \textit{Chain-of-Event}~\cite{yao2024chain} specifically leverage human feedback to refine the causal graphs generated by automated systems, allowing experts to correct erroneous edges and confirm valid causal paths, thereby improving the accuracy of the final diagnosis.
\textit{TraceDiag}~\cite{ding2023tracediag} learns from historical expert decisions through reinforcement learning, capturing pruning policies derived from experienced engineers without requiring real-time expert input.
This balance between automation and expert oversight is critical, as purely manual configuration creates operational overhead while purely automated systems may lack organization-specific context.

A critical aspect of actionability is helping operators understand when to trust automated recommendations.
\textit{Nezha}~\cite{yu2023nezha} provides ranked suspicion lists with confidence measures, allowing experts to apply their judgment rather than blindly following system outputs.
\textit{LM-PACE}~\cite{zhang2024lm} implements a sophisticated two-stage prompting approach that quantifies evidence strength from historical incidents and provides calibrated confidence scores.
These mechanisms enable more informed decision-making about when to follow automated guidance versus seeking additional expert input, though the challenge remains in developing confidence metrics that accurately reflect system reliability while remaining interpretable under operational pressure.

Interactive visualization interfaces further enhance actionability by presenting complex dependency relationships in comprehensible formats.
Systems such as \textit{GMTA}~\cite{guo2020graph}, \textit{Eadro}~\cite{lee2023eadro}, and \textit{Groot}~\cite{wang2021groot} provide user-friendly interfaces that help operators quickly validate findings and identify remediation actions.
The design challenge lies in presenting sufficient detail for expert validation while maintaining clarity and usability during high-stress incident response.

\section{RESEARCH TREND AND DISTRIBUTIONS} \label{sec:trend}

In this section, we analyze the research trend and distributions of root cause analysis through the lens of our seven-goal taxonomy (defined in Section~\ref{sec:preliminaries}).
We examine how the field has evolved in terms of publication volume, research focus, and the diversity of problem settings addressed by the community.

\subsection{Publication Trend and Research Focus Evolution}

\begin{figure}[t]
    \centering
    \includegraphics[width=0.7\linewidth]{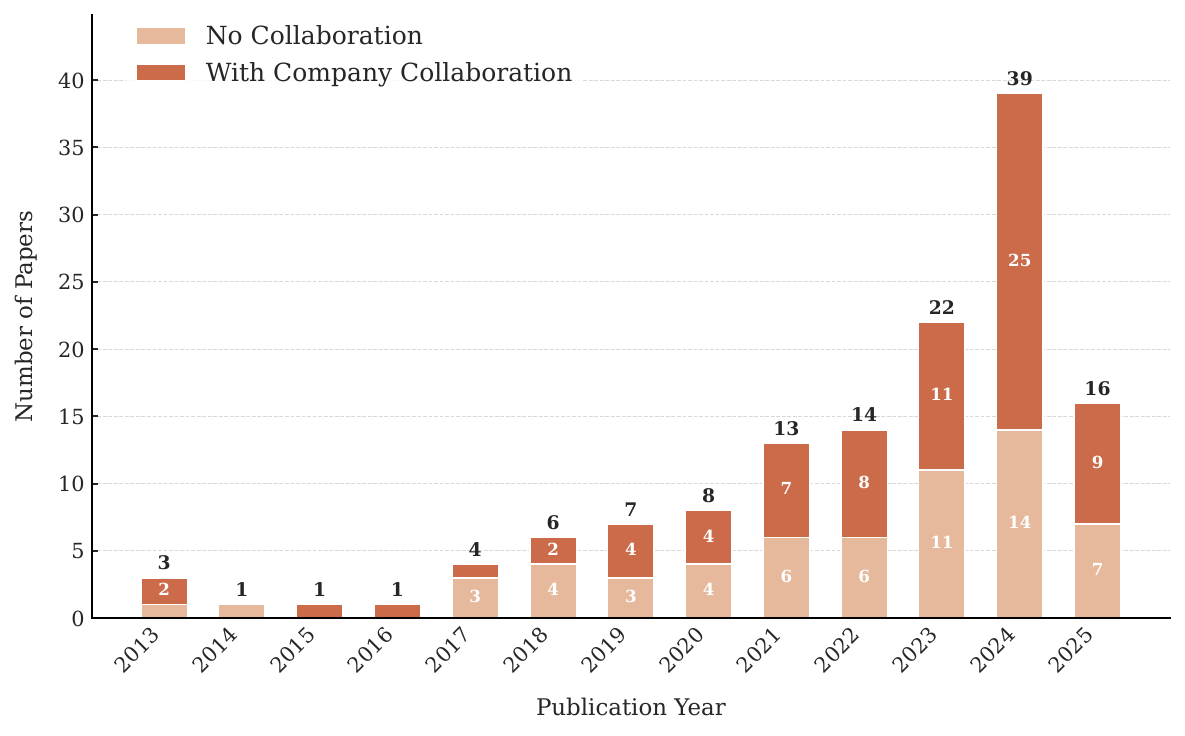}
    \caption{The number of collected papers published per year in top-tier venues, distinguishing between those with and without company collaboration.}
    \label{fig:papertrend}
\end{figure}

We first analyze the publication trend of RCA research in top-tier venues.
As shown in Fig.~\ref{fig:papertrend}, the data reveal a clear upward trend in the number of publications over the past decade.
Particularly, the field has experienced significant growth since 2021, with the number of papers peaking at 40 in 2024.
This steady increase underscores the growing importance and recognition of RCA research within the academic community.
Furthermore, the figure highlights a substantial rise in papers involving collaboration with industry, especially in recent years.
This trend suggests that RCA research is not only gaining academic interest but is also increasingly addressing challenges of significant practical relevance to the industry.

\begin{figure}[t]
    \centering
    \includegraphics[width=1\linewidth]{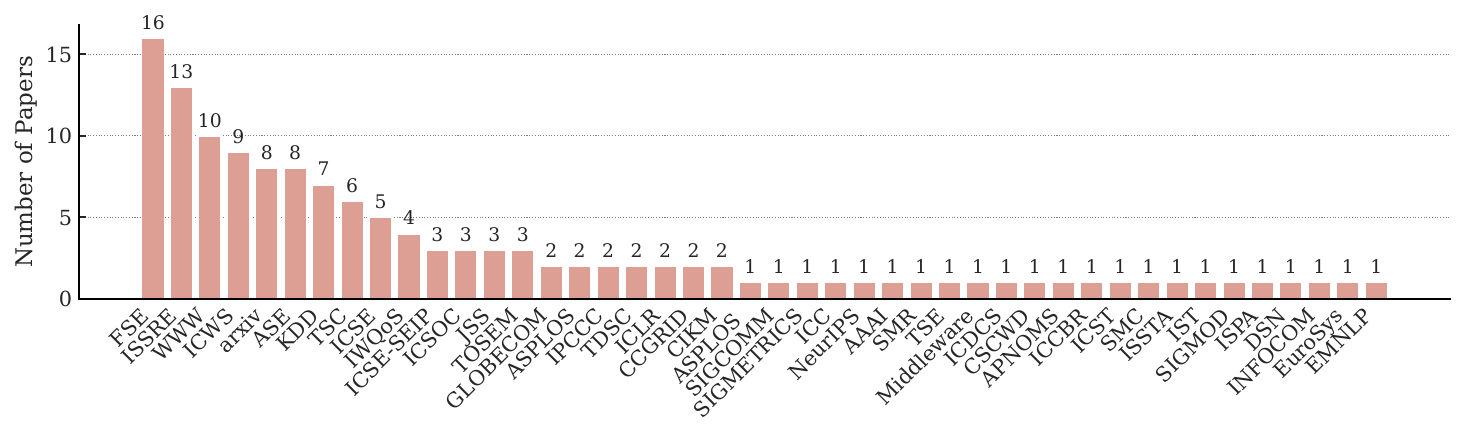}
    \caption{Distribution of the collected papers across various research venues.}
    \label{fig:conferencetrend}
\end{figure}

Fig.~\ref{fig:conferencetrend} illustrates the distribution of these papers across a diverse set of prestigious research venues.
A significant portion of RCA papers are published in premier software engineering venues, such as FSE, ISSRE, ASE, and ICSE.
At the same time, there are notable contributions in leading data mining (e.g., KDD), systems (e.g., ASPLOS), and security (e.g., TDSC) venues.
This broad distribution highlights that RCA is a topic of interest that spans multiple research fields, demonstrating its wide-ranging applicability and relevance.

\subsection{Settings of RCA}

\begin{figure}[t]
    \centering
    \includegraphics[width=0.9\linewidth]{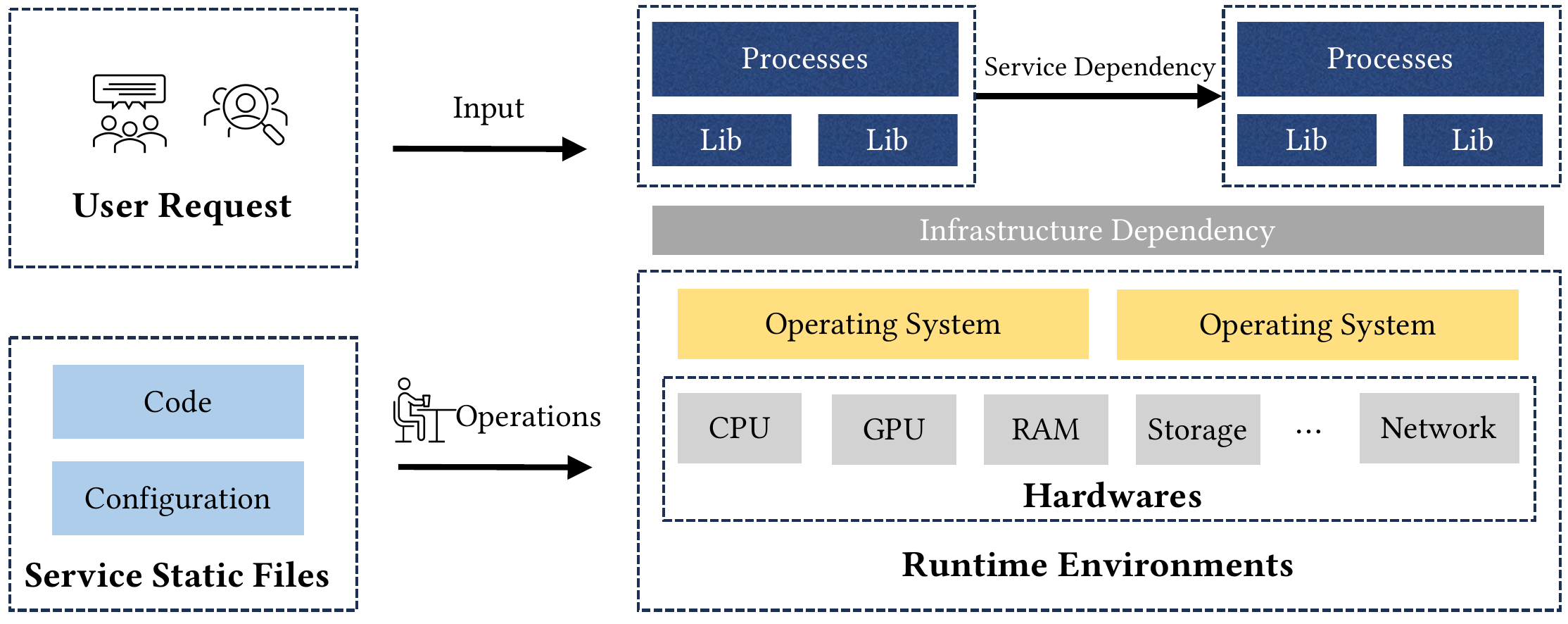}
    \caption{Hierarchical Root Cause Model, showing all potential locations where a root cause can manifest in a microservice system.}
    \label{fig:rootcausemodel}
\end{figure}

The diversity of ground truth root causes directly impacts the effectiveness of RCA models.
In practice, failures can originate from any component in the root cause runtime model shown in Fig.~\ref{fig:rootcausemodel}.
By analyzing \totalrcapaper papers that explicitly discuss ground truth root causes used in their experiments, we categorize these failures into six groups based on their point of origin:

\begin{itemize}
\item \textit{Resource}: Failures originating from the exhaustion or contention of fundamental system resources, where the system's provisioned capacity is insufficient for the given workload.
This category includes issues like CPU saturation, memory exhaustion (OOM), insufficient disk space, I/O bottlenecks, and network bandwidth limitations, assuming no underlying software defects are the direct cause.
For example, this applies when a correctly implemented service fails due to under-provisioning.
\item  \textit{Code}: Failures originating from defects within the service's source code.
This includes logic errors, concurrency bugs, memory leaks, inefficient algorithms leading to performance degradation, or faulty code changes.
A failure is classified as a code issue even if its symptom is resource exhaustion (e.g., a memory leak causing an OOM error), because the initial trigger is the software defect itself.
\item \textit{Configuration}: Failures originating from incorrect values in externalized application or environment settings.
This includes misconfigured parameters in deployment files (e.g., YAML, Helm charts), incorrect environment variables, or erroneous settings in external configuration services.
This category is distinct from code-level defects as these issues can be remediated without changing the compiled source code.
\item \textit{User Request}: Failures originating from the characteristics of incoming user requests that the system cannot handle correctly.
This includes unexpected request patterns (e.g., traffic surges, DDoS attacks), malformed or malicious inputs (e.g., SQL injection payloads), or ``poison pill'' requests that trigger latent bugs.
\item \textit{Infrastructure Dependency}: Failures originating from the functional correctness or availability of underlying, general-purpose infrastructure services.
This includes database deadlocks, message queue service unavailability, or cache service failures.
This category is distinct from \textit{Resource} issues on the infrastructure's host; it concerns the failure of the service provided by the infrastructure itself.
\item \textit{Service Dependency}: Failures originating from an incorrect or unavailable response from another service within the application's architecture or from a third-party API.
This includes cascading failures where an upstream service fails due to a fault in a downstream dependent service.
The root cause is attributed to the dependent service that first breaks the expected contract.
\end{itemize}

These categories directly map to the hierarchical structure presented in Fig.~\ref{fig:rootcausemodel}.
\textit{Resource} and \textit{Infrastructure Dependency} failures correspond to the foundational infrastructure layer.
\textit{Code} and \textit{Configuration} issues are located within the service implementation layer.
Finally, \textit{User Request} and \textit{Service Dependency} failures manifest at the service interaction layer, representing external and internal dependencies, respectively.
This mapping provides a structured view of how different types of failures are situated within a microservice architecture.

\begin{figure}[t]
    \centering
    \includegraphics[width=0.5\linewidth]{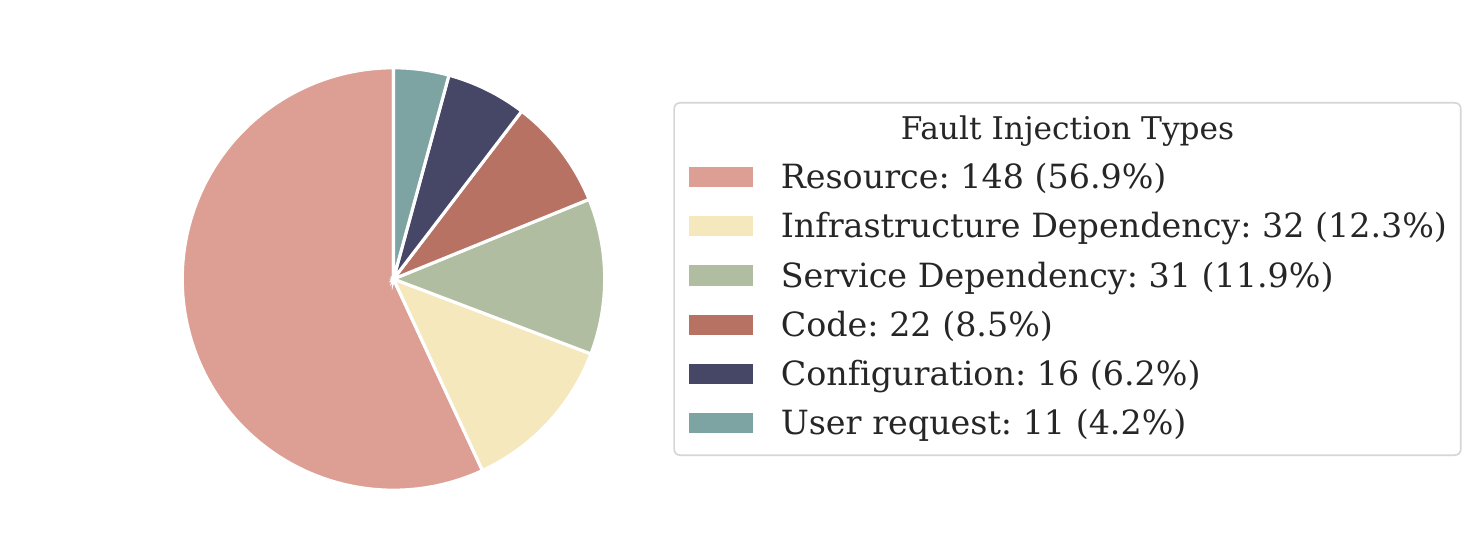}
    \caption{Distribution of root cause types across \totalrcapaper papers. The chart shows the proportion of studies focusing on each root cause type. Note that a single paper may address multiple root cause types and is thus counted in each applicable category.}
    \label{fig:rootcausedistribution}
\end{figure}

As shown in Fig.~\ref{fig:rootcausedistribution}, the high proportion of papers focusing on resource-related failures (56.9\%) indicates that much of the RCA research prioritizes resource management issues.
However, this heavy focus raises important questions about whether RCA research is adequately addressing other critical failure types.
In real-world microservice systems, failures often arise from complex or less obvious causes, such as misconfigurations, code bugs, or intricate interactions between services.
The relatively low percentages for code-related (8.5\%) and configuration-related (6.2\%) failures are particularly noteworthy.
In practice, misconfigurations and software bugs are common sources of severe outages and performance degradation.
This discrepancy suggests that current RCA research may not be adequately addressing these significant real-world challenges.
Additionally, with only 4.2\% of studies focusing on user request-related failures, there is a notable gap in research addressing how user behaviors, both intentional and unintentional, contribute to system failures.

This distribution pattern reflects an implicit simplification strategy in the field.
From the perspective of our formalization in Section~\ref{sec:formal_rca}, resource-related root causes typically produce clear, observable signals in metrics (as part of the observation space $\mathcal{O}$), making them easier to diagnose from a technical perspective.
In contrast, detecting failures from code bugs, misconfigurations, or user requests often requires more sophisticated analysis of logs, traces, and their correlations, which substantially increases the complexity of inference within the problem space.


\subsection{Benchmarks}

\begin{table}[t]
\centering
\footnotesize
\caption{Root Cause Analysis Types and Detailed Entities}
\label{tab:benchmark}
\begin{tabular}{p{2.4cm}|p{1cm}|p{5.5cm}|p{1.2cm}|p{1.5cm}} 
\toprule
\textbf{Benchmark}\tablefootnote{We are only counting the core business logic code and excluding any auto-generated code or infrastructure services like databases.} & 
\textbf{Svc \#} & 
\textbf{LoC/Programming Languages} & 
\textbf{Protocol} & 
\textbf{Last Update}  \\ 
\midrule
TrainTicket~\cite{trainticket} & 45 & 37746/Java, 23/Go, 292/Python, 5335/JavaScript, 9733/HTML & HTTP & 2022-11-01 \\ 
\midrule
Online Boutique~\cite{OnlineBoutique} & 11 & 5881/Go, 1043/Python, 740/HTML, 634/C\#, 347/JavaScript, 255/Java & gRPC & 2024-10-03 \\ 
\midrule
Sock Shop~\cite{sockshop} & 9 & 4010/JavaScript, 3577/Java, 3283/Go, 1640/Python & HTTP & 2023-12-05 \\ 
\midrule
HotelReservation~\cite{DeathStarBench} & 10 & 7298/Go & gRPC & 2024-06-28 \\ 
\midrule
SocialNetwork~\cite{DeathStarBench} & 12 & 5753/C++ & Thrift & 2024-06-28 \\
\midrule
Astronomy Shop~\cite{AstronomyShop} & 14 & 10306/C++, 6714/TypeScript, 4452/Go, 1286/JavaScript, 1045/Elixir, 1035/Python, 696/C\#, 401/Java, 313/Rust, 193/PHP, 67/Kotlin, 53/Ruby & gRPC/HTTP & 2025-10-21 \\
\midrule
TeaStore~\cite{TeaStore} & 7 & 12317/Java, 1693/JavaScript & HTTP & 2025-01-08 \\
\bottomrule
\end{tabular}
\end{table}

This section lists the public benchmarks in the literature based on our collected papers. 
Publicly available benchmarks are important to RCA research since the companies usually cannot open-source their data due to privacy and security issues. 
Additionally, root cause analysis is very related to the industry practice, and the cloud environment is highly dynamic. 
As a result, researchers need publicly available benchmarks to inject corresponding failures, stimulate the environment, and show the effectiveness of the proposed method.

Table~\ref{tab:benchmark} provides a detailed summary of seven prominent public benchmarks used in RCA research.
The table details each benchmark's scale (number of services and lines of code), technological stack (programming languages and protocols), and maintenance status (last update).
The benchmarks exhibit significant diversity.
TrainTicket~\cite{trainticket} remains the largest in terms of service count, while TeaStore~\cite{TeaStore} and Astronomy Shop~\cite{AstronomyShop} feature substantial Java and polyglot codebases, respectively.
Protocols range from the conventional HTTP to the more modern gRPC and Thrift, reflecting varied architectural choices in microservice systems.

However, a critical analysis of their maintenance status reveals a significant challenge for the research community.
Several foundational benchmarks show signs of abandonment.
TrainTicket~\cite{trainticket} has not been updated in over two years, and Sock Shop~\cite{sockshop} is officially archived.
DeathStarBench~\cite{DeathStarBench} also suffers from stalled development, with only a subset of its promised benchmarks released and recent activity confined to minor bug fixes.
In contrast, Online Boutique~\cite{OnlineBoutique}, TeaStore~\cite{TeaStore}, and the newly introduced Astronomy Shop~\cite{AstronomyShop} appear to be under active maintenance, receiving regular updates.
This maintenance gap creates a bifurcation in the landscape, where newer research may gravitate towards actively maintained but potentially less complex systems.

A more profound limitation lies in the inherent design of these benchmarks.
While they provide environments for evaluating fault localization, their static nature and limited complexity hinder research on higher-level RCA goals~\cite{fang2025empirical}.
For instance, they are ill-suited for studying \textit{Adaptive Learning}, as they do not model system evolution.
Furthermore, their fault injection capabilities are often restricted to resource-level failures, offering little support for investigating complex, non-resource-related issues like logic bugs or misconfigurations.
This deficiency makes it difficult to develop and validate methods aimed at achieving deep \textit{Interpretability} or \textit{Actionability}.
This maintenance and complexity bottleneck suggests that the community needs either renewed investment in existing benchmarks or a shift toward automatically-generated, dynamically-maintained benchmark environments.

\subsection{Datasets}

\begin{table*}[t]
\centering
\footnotesize
\caption{Public dataset for root cause analysis (M for metrics, L for logs, and T for traces)}
\label{tab:dataset}
\begin{tblr}{
  cells = {c},
  vline{5} = {-}{},
  hline{1-2,20} = {-}{},
  row{1} = {font=\bfseries},
}
Dataset & Type & Format & Amount & Dataset & Type & Format & Amount \\
Dycause\cite{dycausedataset} & M & XLSX & 4.1MB & DéjàVu-A1\cite{dejavudataset} & M & CSV & 75.1MB \\
GrayScope\cite{grayscopedataset} & M & CSV & 8.4MB & DéjàVu-A2\cite{dejavudataset} & M & CSV & 82.2MB \\
RCD-SS\cite{rcddataset} & M & CSV & 16MB & MicroCU\cite{microcudataset} & M & npy & 118MB \\
DéjàVu-C\cite{dejavudataset} & M & CSV & 48.8MB & RCAEval1-OB\cite{barodataset} & M & CSV & 166.84MB \\
ChangeRCA-OB\cite{changercadataset} & M & CSV & 60MB & RCAEval1-SS\cite{barodataset} & M & CSV & 484.17MB \\
DéjàVu-B\cite{dejavudataset} & M & CSV & 1.7GB & RCAEval1-TT\cite{barodataset} & M & CSV & 1.75GB \\
LatentScope\cite{latentscopedataset} & M & JSON & 2.1GB & DéjàVu-D\cite{dejavudataset} & M & CSV & 3.7GB \\
Squeeze\cite{squeezedataset} & M & CSV & 18GB & MEPFL-SS\cite{mepfldataset} & T & CSV & 59MB \\
MEPFL-TT\cite{mepfldataset} & T & CSV & 2.3GB & AIOps Comp-2020\cite{aiopsdataset} & M, T & CSV & 16GB \\
Murphy-DSB\cite{murphydataset} & M, T & JSON & 99GB & GAMMA\cite{gammadataset} & M, L & RAW format/CSV & 39GB \\
RCAEval2-SS\cite{RCAEval} & M, L & CSV & 2.16GB & RCAEval3-SS\cite{RCAEval} & M, L & CSV & 872.18MB \\
Nezha-TT\cite{nezhadataset} & M, T, L & CSV & 351MB & Eadro-TT\cite{eadrodataset} & M, T, L & CSV/JSON & 841MB \\
FAMOS-TT\cite{duan2025famos} & M, T, L & CSV/Parquet & 925MB & Eadro-SN\cite{eadrodataset} & M, T, L & CSV/JSON & 1.3GB \\
RCAEval3-OB\cite{RCAEval} & M, T, L & CSV & 1.31GB & RCAEval3-TT\cite{RCAEval} & M, T, L & CSV & 2.07GB \\
Nezha-OB\cite{nezhadataset} & M, T, L & CSV & 2.5GB & AIOps Comp-2025\cite{aiopsdataset25} & M, T, L & Parquet & 5.8GB \\
RCAEval2-OB\cite{RCAEval} & M, T, L & CSV & 8.05GB & FAMOS-Mall\cite{duan2025famos} & M, T, L & CSV/Parquet & 9.55GB \\
Fang et al.\cite{fang2025empirical} & M, T, L & Parquet & 12.70GB & RCAEval2-TT\cite{RCAEval} & M, T, L & CSV & 21.72GB \\
AIOps Comp-2021\cite{aiopsdataset} & M, T, L & CSV & 25GB & GAIA\cite{gaiadataset} & M, T, L & CSV & 41GB \\
\end{tblr}
\end{table*}

This section presents an overview of the publicly available datasets relevant to root cause analysis (RCA), as identified from the literature. 
Table~\ref{tab:dataset} summarizes the key attributes of these datasets, including their data types (metrics, traces, logs), data formats (\eg CSV, JSON), dataset size (calculated after decompression), and the research papers that utilized these datasets.
Most datasets are collected from the public benchmarks, \eg Online Boutique~\cite{OnlineBoutique} (OB), SockShop~\cite{sockshop} (SS), TrainTicket~\cite{trainticket} (TT), DeathStarBench~\cite{DeathStarBench} (DSB), SocialNetwork~\cite{DeathStarBench} (SN).
The table reveals a clear trend toward multi-modal and larger-scale datasets over time.
Early datasets like Dycause~\cite{dycausedataset} and GrayScope~\cite{grayscopedataset} were small and focused exclusively on metrics (M).
In contrast, more recent contributions such as GAIA~\cite{gaiadataset}, FAMOS~\cite{duan2025famos}, and the RCAEval series~\cite{RCAEval} incorporate a combination of metrics (M), logs (L), and traces (T), with sizes scaling into tens of gigabytes.
This evolution reflects the community's growing recognition that effective RCA requires a holistic view of the system, integrating diverse telemetry sources.

Despite this progress, a critical gap persists from the perspective of our formal framework (Section~\ref{sec:formal_rca}).
While these datasets provide rich subsets of the observation space $\mathcal{O}$, nearly all of them lack ground truth labels for the complete incident propagation graph $\mathcal{G}$.
The provided labels are typically confined to identifying the root cause node, such as a faulty service or metric.
This limitation reinforces a "point-finding" research paradigm, where the primary goal is to pinpoint a single fault origin.
This scarcity of comprehensively labeled, graph-based ground truth is a major bottleneck for developing and evaluating methods aimed at achieving higher-level goals like \textit{Interpretability}.
Without the ground truth of the causal chain, researchers cannot validate the correctness of inferred causal paths, hindering the transition toward more sophisticated "graph-building" RCA models.
Future dataset collection efforts must prioritize capturing not just the root cause, but the entire causal chain of events to foster this next generation of RCA research.

\subsection{Public Available Tools}

\begin{table*}[t]
\centering
\caption{Publicly Available Tools/Codes for Root Cause Analysis}
\label{tab:publicartifacts}
\resizebox{\textwidth}{!}{%
\begin{tblr}{
  cells = {c},
  cell{1}{1,4,7,10,13} = {font=\bfseries},
  vline{4,7,10,13} = {-}{},
  hline{1,12} = {-}{0.08em},
  hline{2} = {-}{},
  row{1} = {font=\bfseries},
}
Tool & URL & Year & Tool & URL & Year & Tool & URL & Year & Tool & URL & Year & Tool & URL & Year \\
Sieve & \cite{sieve} & 2017 & Log3C & \cite{Log3C} & 2018 & Squeeze & \cite{Squeeze} & 2019 & MEPFL & \cite{MEPFL} & 2019 & TraceAnomaly & \cite{TraceAnomaly} & 2020 \\
MicroRank & \cite{MicroRank} & 2021 & TraceRCA & \cite{TraceRCA} & 2021 & DyCause & \cite{DyCause} & 2021 & PDiagnose & \cite{PDiagnose} & 2021 & DéjàVu & \cite{DejaVu} & 2022 \\
GIED & \cite{GIED} & 2022 & CIRCA & \cite{CIRCA} & 2022 & MicroCBR & \cite{MicroCBR} & 2022 & RCD & \cite{rcd} & 2022 & SwissLog & \cite{SwissLog} & 2022 \\
Diagfusion & \cite{DiagFusion} & 2023 & MicroCU & \cite{MicroCU} & 2023 & CMDiagnostor & \cite{CMDiagnostor} & 2023 & CausalRCA & \cite{CausalRCA} & 2023 & Nezha & \cite{Nezha} & 2023 \\
Eadro & \cite{Eadro} & 2023 & ShapleyIQ & \cite{ShapleyIQ} & 2023 & TraceStream & \cite{TraceStream} & 2023 & BARO & \cite{baro} & 2024 & LoFI & \cite{LoFI} & 2024 \\
ART & \cite{ART} & 2024 & LatentScope & \cite{LatentScope} & 2024 & HeMiRCA & \cite{HeMiRCA} & 2024 & Chain-of-Event & \cite{ChainofEvent} & 2024 & DeepHunt & \cite{DeepHunt} & 2024 \\
ChangeRCA & \cite{ChangeRCA} & 2024 & KPIRoot & \cite{KPIRoot} & 2024 & CHASE & \cite{CHASE} & 2024 & Medicine & \cite{Medicine} & 2024 & MicroIRC & \cite{MicroIRC} & 2024 \\
UniDiag & \cite{UniDiag} & 2024 & MicroDig & \cite{MicroDig} & 2024 & RUN & \cite{RUN} & 2024 & SLIM & \cite{SLIM} & 2024 & mABC & \cite{mABC} & 2024 \\
LasRCA & \cite{LasRCA} & 2024 & FaaSRCA & \cite{FaaSRCA} & 2024 & SCELM & \cite{SCELM} & 2025 & DiagMLP & \cite{DiagMLP} & 2025 & LEMMA-RCA & \cite{LEMMARCAREF} & 2025 \\
RCAEval & \cite{RCAEval} & 2025 & AERCA & \cite{AERCA} & 2025 & TVdiag & \cite{TVdiag} & 2025 & TrioXpert & \cite{TrioXpert} & 2025 & FAMOS & \cite{FAMOS} & 2025 \\
\end{tblr}%
}
\end{table*}

This section compiles a collection of publicly accessible toolkits and codebases that can facilitate further research in root cause analysis. 
Among the ~\totalrcapaper papers we reviewed, 50 have made their implementations publicly available, as cataloged in Table~\ref{tab:publicartifacts}.
The chronological listing of these artifacts reveals a significant acceleration in open-source contributions, with the number of tools released in 2024 alone nearly matching the total from all preceding years combined.
This trend toward openness, exemplified by recent and notable tools like BARO~\cite{pham2024baro}, LatentScope~\cite{xie2024microservice}, and Chain-of-Event~\cite{yao2024chain}, is crucial for promoting transparency, reproducibility, and standardized evaluation within the community.

The evolution of these tools also mirrors the field's methodological shifts.
Early contributions such as Sieve~\cite{thalheim2017sieve} and Log3C~\cite{Log3C} laid the groundwork, while more recent systems like Eadro~\cite{lee2023eadro} and FAMOS~\cite{FAMOS} demonstrate increasing sophistication in handling multi-modal data and complex causal inference.
The availability of this diverse array of open-source tools provides an invaluable resource for the community.
It enables researchers to benchmark new techniques against established baselines, adapt existing models for novel scenarios, and build upon prior work to push the frontiers of the field.
This collaborative ecosystem is essential for systematically advancing RCA research.

\section{DISCUSSION}\label{sec:discussion}

\subsection{Threats to Validity}

\textit{Data source credibility.} 
This survey covers only a subset of the available literature, with a focus on papers related to microservices root cause analysis published in top-tier conferences and journals over the past decade. 
Due to limitations in both time and resources, it was not feasible to collect all relevant works, which may result in some incompleteness. 
For instance, while the pipeline in RootCLAM~\cite{han2023root} aligns with the general scope of this paper, which encompasses anomaly detection, root cause localization, and anomaly mitigation, the specific context of RootCLAM~\cite{han2023root} is quite different from ours. 
RootCLAM~\cite{han2023root} utilizes a loan approval scenario based on the German Credit dataset, which falls outside the domain of incident management. 
Consequently, works that do not pertain to incident management, such as RootCLAM, were excluded from our discussion. 
However, its inclusion highlights the broader applicability of root cause analysis, which extends beyond incident management scenarios.

Moreover, while we have strived to ensure the accuracy of our literature understanding and analysis, there is an inherent risk of subjective interpretation errors during the reading process. To mitigate these risks, we employed a cross-validation approach: the primary authors independently read and summarized the papers, followed by a cross-review of the results to enhance the accuracy and consistency of our findings.

\subsection{Implications of the Goal-Driven Framework}

Our seven-goal framework serves not merely as a taxonomy, but as an analytical lens to reveal the underlying dynamics of RCA research.
The goals are not independent; they exist in a web of synergies and tensions that shape the field.

\textit{Synergies and Tensions.}
There are clear synergistic relationships.
For instance, advances in \textit{Interpretability} (Goal 5), particularly the construction of causal graphs, directly enhance \textit{Actionability} (Goal 7) by providing a logical basis for automated remediation.
Conversely, fundamental tensions force trade-offs.
The pursuit of \textit{Real-time Performance} (Goal 4) often necessitates algorithmic shortcuts that may compromise the depth of \textit{Interpretability} (Goal 5).
Similarly, highly \textit{Adaptive} models (Goal 3) may struggle to maintain \textit{Robustness} (Goal 2) against noisy data, highlighting a conflict between dynamic adaptation and static resilience.
Recognizing these trade-offs is crucial for both researchers designing new methods and practitioners selecting tools for specific operational contexts.

\textit{Evolutionary Trends.}
The framework also illuminates an evolutionary trend in the field.
Early research predominantly focused on foundational goals like achieving \textit{Real-time Performance} and \textit{Multi-granularity}.
With the advent of Large Language Models (LLMs), the research frontier is rapidly expanding towards more semantic and human-centric goals, namely \textit{Interpretability} and \textit{Actionability}, signaling a shift from "what happened" to "why it happened and what to do about it."

\subsection{Bridging the Gap to Ideal RCA}

Our formalization of ideal RCA as a function $\mathcal{F}: \mathcal{O} \rightarrow \mathcal{G}$ that maps observations to a complete incident propagation graph, provides a "north star" for the field.
However, our survey reveals a significant gap between this ideal and the current state of the art.

\textit{The Paradigm Shift: From Pinpointing Causes to Graphing Propagation.}
The vast majority of existing research still operates under a simplified paradigm: pinpointing a single root cause node ($r \in \mathcal{G}$).
The true challenge, and the next frontier, lies in the paradigm shift towards constructing the full propagation graph $\mathcal{G}$.
This shift is hindered by three fundamental gaps:
\begin{itemize}
    \item \textbf{The Evaluation Gap:} There is a critical lack of standardized benchmarks and metrics to evaluate the correctness of a generated propagation graph. Without a target to aim for, progress is inherently limited.
    \item \textbf{The Data Gap:} Correspondingly, few, if any, public datasets provide ground-truth propagation graphs for incidents, making supervised learning of such structures nearly impossible.
    \item \textbf{The Methodology Gap:} Current methods are not explicitly optimized for graph generation. Causal discovery algorithms struggle with the scale and dynamics of real systems, while LLMs, despite their semantic power, lack the structural grounding to reliably produce verifiable causal chains.
\end{itemize}

\textit{Revisiting Triggers vs. Root Causes.}
Furthermore, the ideal graph $\mathcal{G}$ must distinguish between the underlying \textit{root cause} (e.g., a buggy code commit) and the \textit{trigger} (e.g., a specific user input that activates the bug).
This distinction, vital for both immediate mitigation and long-term fixes, is largely overlooked in current research but is essential for achieving a truly comprehensive diagnosis.

\subsection{Future Research Frontiers}

Based on the identified gaps and implications, we outline three forward-looking research frontiers that aim to systematically advance the field towards the ideal of RCA.

\textit{The Next-Generation RCA Benchmark.}
The most pressing need is a community-driven effort to build a large-scale, multi-modal benchmark dataset where incidents are annotated not with single root causes, but with complete, ground-truth propagation graphs ($\mathcal{G}$).
Such a benchmark would catalyze the development and rigorous evaluation of the next generation of RCA models.

\textit{Unified Models for Causal Graph Generation.}
Future research should move beyond single-cause localization and focus on novel architectures designed explicitly for end-to-end propagation graph generation.
This may involve hybrid models that fuse the structural reasoning of GNNs, the semantic and common-sense understanding of LLMs, and the exploratory capabilities of reinforcement learning to navigate vast diagnostic search spaces.

\textit{Deep Integration with the Software Engineering Lifecycle.}
The output of RCA should not be the end of the story.
A truly impactful direction is to create a feedback loop from RCA back into the software engineering lifecycle.
The generated graph $\mathcal{G}$ could automatically inform and trigger actions such as pinpointing the exact faulty code commit, generating new regression tests that codify the failure scenario, and providing actionable insights for architectural redesign, thus transforming RCA from a reactive diagnostic tool into a proactive driver of system reliability.

\section{RELATED WORK}\label{sec:relatedwork}

Several surveys have provided valuable overviews of the Root Cause Analysis (RCA) landscape.
However, their taxonomies are often based on surface-level features, such as input data modalities (\eg logs, metrics, traces).
We argue that such classifications obscure the underlying strategic goals that drive RCA research.
Our work distinguishes itself by introducing a goal-driven framework that offers a more insightful and functionally relevant perspective.

\paragraph{Surveys Based on Data Modality and Techniques}
The most related surveys from Soldani et al.~\cite{soldani2022anomaly}, Zhang et al.~\cite{zhang2024failure}, and Wang et al.~\cite{wang2024comprehensive} organize the field primarily by data sources.
While providing a useful catalog of methods, this data-centric view has a key limitation: the mapping between data types and research objectives is not one-to-one.
For example, methods using different data types might share the same goal (\eg \textit{Real-time Performance}), while methods using the same data type might pursue different goals (\eg \textit{Interpretability} vs. \textit{Adaptive Learning}).
This hinders a clear comparison of methodological trade-offs.
Our survey moves beyond this by classifying works based on the seven fundamental goals defined in Section~\ref{sec:goals_definition}.

\paragraph{Surveys on Specific RCA Sub-domains}
Other works focus on narrower aspects of RCA, such as causal inference-based methods~\cite{pham2024root} or fault localization in software engineering~\cite{wong2016survey}.
These specialized surveys are complementary to our work.
Our goal-driven framework can contextualize their findings within the broader RCA landscape.
For instance, challenges in causal graph construction~\cite{pham2024root} directly relate to the pursuit of \textit{Interpretability} (Goal 5) and the gaps we identify in Section~\ref{sec:discussion}.

\paragraph{Our Unique Contribution}
Unlike previous surveys that document "what has been done," our work provides a conceptual framework to understand "why it was done" and "where to go next."
By focusing on seven orthogonal goals, we enable a more meaningful comparison of approaches and illuminate the inherent trade-offs in the field.
Furthermore, our formalization of the ideal RCA output as an incident propagation graph ($\mathcal{G}$) provides a "north star" for the community, allowing us to systematically identify the gap between the current state of the art and the ultimate goal of explaining the full causal story of an incident.
In summary, our survey offers a more profound, goal-oriented synthesis that provides a clearer roadmap for future research.

\section{CONCLUSION}\label{sec:conclusion}

In this paper, we addressed the fragmentation that has long characterized Root Cause Analysis (RCA) research, a field where the prevalence of task-specific solutions has hindered a unified understanding.
We argued that traditional surveys, which categorize methods by data types, fail to capture the underlying objectives driving the research.
In response, we proposed a new conceptual framework built on two core contributions: a formal definition of ideal RCA as the generation of a complete incident propagation graph ($\mathcal{G}$), and a seven-goal taxonomy derived from the practical needs of the incident management lifecycle.
Through the lens of this goal-driven framework, our analysis systematically deconstructed the field.
This revealed the inherent synergies and trade-offs, such as the tension between real-time performance and interpretability, that shape all RCA systems.
More critically, this perspective allowed us to identify a fundamental gap between the community's dominant focus on pinpointing single root causes and the ideal of constructing the full propagation graph.
We contend that bridging this gap is the central challenge for the field, requiring a focused effort to overcome fundamental obstacles in evaluation, data availability, and methodology.
Building on these insights, we outlined a future research agenda centered on three key areas: creating a next-generation benchmark with ground-truth graphs, developing unified models for causal graph generation, and deeply integrating RCA into the software engineering lifecycle.
Ultimately, this survey aims not only to document the past but also to provide a structured roadmap to guide the community toward a more systematic and impactful future, advancing root cause analysis from a reactive practice into a proactive, data-driven science.



\appendix

\bibliographystyle{ACM-Reference-Format}
\bibliography{ref}

\end{document}